\documentclass[12pt]{article}
\usepackage{graphicx}
\usepackage{amsmath}
\usepackage{amssymb}
\usepackage{subfigure}
\usepackage{latexsym}
\usepackage{hyperref}
\usepackage[sort]{cite}

{\catcode`\|=\active 
  \gdef\Braket#1{\left<\mathcode`\|"8000\let|\bravert {#1}\right>}}
\def\bravert{\egroup\,\vrule\,\bgroup}
\newcommand{\at}{{\tilde a}}

\newcommand{\be}{\begin{eqnarray}}
\newcommand{\ee}{\end{eqnarray}}
\newcommand{\bea}{\begin{eqnarray}}
\newcommand{\eea}{\end{eqnarray}}
\newcommand{\ben}{\begin{equation}}

\newcommand{\nn}{\nonumber}

\numberwithin{equation}{section}

\newsavebox{\ns}
\newsavebox{\dbrane}
\newsavebox{\dbshort}

\usepackage{epsfig}
\usepackage{amssymb}

\def\appendix{{\newpage\section*{Appendix}}\let\appendix\section%
        {\setcounter{section}{0}
        \gdef\thesection{\Alph{section}}}\section}

\newcommand\ba{\begin{eqnarray}}
\newcommand\ea{\end{eqnarray}}

\def\Dslash{\,\,{\raise.15ex\hbox{/}\mkern-12mu D}}
\def\Dbarslash{\,\,{\raise.15ex\hbox{/}\mkern-12mu {\bar D}}}
\def\delslash{\,\,{\raise.15ex\hbox{/}\mkern-9mu \partial}}
\def\delbarslash{\,\,{\raise.15ex\hbox{/}\mkern-9mu {\bar\partial}}}
\def\pslash{\,\,{\raise.15ex\hbox{/}\mkern-10mu p}}
\def\vslash{\,\,{\raise.15ex\hbox{/}\mkern-9mu V}}
\def\kslash{\,\,{\raise.15ex\hbox{/}\mkern-9mu k}}
\def\calDslash{\,\,{\raise.15ex\hbox{/}\mkern-12mu {\cal D}}}

\newcommand{\uu}{^}
\newcommand{\pp}{\partial}
\renewcommand{\L}{\Lambda}
\renewcommand{\exp}[1]{{\rm exp}\left( #1 \right)}

\newcommand{\m}{\mu}

\renewcommand{\m}{\mu}
\newcommand{\n}{\nu}
\newcommand{\s}{\sigma}

\newcommand{\G}{\Gamma}

\renewcommand{\o}{\omega}

\newcommand{\sqd}{^2}

\renewcommand{\b}{\beta}

\newcommand{\apr}{{\alpha^\prime} {}}

\newcommand{\IZ}{\relax\ifmmode\mathchoice
{\hbox{\cmss Z\kern-.4em Z}}{\hbox{\cmss Z\kern-.4em Z}}
{\lower.9pt\hbox{\cmsss Z\kern-.4em Z}} {\lower1.2pt\hbox{\cmsss
Z\kern-.4em Z}}\else{\cmss Z\kern-.4em Z}\fi} \font\cmss=cmss10
\font\cmsss=cmss10 at 7pt
\newcommand{\inbar}{\,\vrule height1.5ex width.4pt depth0pt}
\newcommand{\IC}{{\relax\hbox{$\inbar\kern-.3em{\rm C}$}}}
\newcommand{\IQ}{{\relax\hbox{$\inbar\kern-.3em{\rm Q}$}}}
\newcommand{\IP}{\relax{\rm I\kern-.18em P}}
\newcommand{\Ione}{{\relax\hbox{$\inbar\kern-.39em{\rm 1}$}}}

\newcommand{\ed}{\dot{e}}

\renewcommand{\o}{\omega}

\newcommand{\ct}{\tilde{c}}

\renewcommand{\L}{\Lambda}

\renewcommand{\at}{{\tilde{\alpha}}}

\newcommand{\IR}{\relax{\rm I\kern-.18em R}}
\def\blfootnote{\xdef\@thefnmark{}\@footnotetext}

\newcommand{\bm}{\begin{matrix}}

\newcommand{\rr}[1]{(\ref{{#1}})}

\newcommand{\xxx}{(\xx)}

\newcommand{\heading}[1]{\ \\ \noindent {\bf #1}\\ }

\def\bi{\begin{itemize}}
\def\ei{\end{itemize}}

\def\ed{\end{document}}

\renewcommand{\rr}[1]{(\ref{#1})}

\def\ct{{\cal T}}

\renewcommand\xxx{\nn\\ &&\ \nn\\  }

\newcommand{\stopdoc}{\cheatsheet \end{document}}

\def\at{{\tilde a}}
\begin{document}

\begin{titlepage}
\begin{flushright}
arXiv:0804.2262 [hep-th]
\end{flushright}
\vspace{15 mm}
\begin{center}
{\Large \bf  Cosmology of the closed string tachyon }
\end{center}
\vspace{6 mm}
\begin{center}
{ Ian Swanson }\\
\vspace{6mm}
{\it School of Natural Sciences, Institute for Advanced Study\\
Princeton, NJ 08540, USA }
\end{center}
\vspace{6 mm}
\begin{center}
{\large Abstract}
\end{center}
\noindent
The spacetime physics of bulk closed 
string tachyon condensation is studied at the level of a two-derivative effective action.
We derive the unique perturbative tachyon potential consistent with a full class
of linearized tachyonic deformations of supercritical string theory.  The solutions 
of interest deform a general linear dilaton background by the
insertion of purely exponential tachyon vertex operators.  
In spacetime, the evolution of the tachyon drives an 
accelerated contraction of the universe and,
absent higher-order corrections, the theory collapses to a cosmological singularity in 
finite time, at arbitrarily weak string coupling.   
When the tachyon exhibits a null symmetry, the worldsheet dynamics are known to be exact 
and well-defined at tree level.  We prove that if the two-derivative effective action is free 
of {\it non}-gravitational singularities,
higher-order corrections always resolve the spacetime curvature singularity of the null
tachyon.  The resulting theory provides an explicit mechanism by which tachyon 
condensation can generate or terminate the flow of cosmological time in string theory.
Additional particular solutions can resolve an initial singularity with a tachyonic phase
at weak coupling, or yield solitonic configurations that localize the universe along 
spatial directions.
\vspace{1cm}
\begin{flushleft}
April 15, 2008
\end{flushleft}
\end{titlepage}
\tableofcontents
\newpage

\section{Introduction}
The study of tachyon condensation in open string theory has led to a number of 
important insights into the nature of instabilities in quantum gravity.
Direct calculations in open string field theory have provided detailed evidence in support 
of Sen's conjecture (for useful reviews of this subject, see, 
e.g.,~\cite{Sen:2004nf,Taylor:2003gn,DeSmet:2001af,Ohmori:2001am}, and 
\cite{martinsimeon} for recent developments).  
Namely, the open string tachyon represents an unstable mode of 
a space-filling D-brane.  The process of tachyon condensation drives the decay of the
D-brane, and the endpoint is an excited state of the closed string vacuum that carries the 
energy of the original D-brane.  Solitonic configurations can also arise, represented  
as D-branes filling a lower number of spatial dimensions.  

Attempts to understand closed string tachyon condensation initially focused 
on localized tachyons 
(see, e.g.,~\cite{local1,local2,local3,local4,local5,local6,local7,local8,local9}).
A well-known example is the theory of winding tachyons localized on the conical orbifold
${\mathbb C}/{\mathbb Z}_N$ \cite{local1}.  Studies using brane probes, RG flow and 
string field theory have provided evidence that tachyon condensation in these systems drives 
a reduction in the orbifold rank $N$, and a resolution of the conical singularity.  

Similar to the open string tachyon of a space-filling D-brane,
the {\it bulk} closed string tachyon fills spacetime completely, and presents a 
tantalizing analogy.  Adopting the lessons of 
the open string problem, it is natural to guess that the bosonic bulk closed string 
tachyon signals an instability of spacetime itself, and the transition to a 
stable endpoint of tachyon condensation represents the decay of spacetime altogether.  
According to this analogy, solitons of the tachyon condensate appear as lower 
dimensional spacetime, allowing for a dynamical transition between physical theories 
with different numbers of spatial dimensions.  A number of recent studies have provided 
indirect evidence in support of this picture in bosonic supercritical string 
theory \cite{paper0,paper1,paper2}.\footnote{Corresponding systems in superstring theory 
exhibit a more baroque landscape of stable and semi-stable endpoints of tachyon condensation
\cite{paper4}.}

Overall, the problem of understanding bulk closed string tachyon condensation in detail has 
been approached in roughly three regimes: 
1) At the level of string field theory,
2) from the perspective of the worldsheet CFT, and
3) within the framework of spacetime effective theories.

\heading{String field theory}
Relative to the corresponding problem in open string theory, attempts to study closed string
tachyon condensation using field theoretic techniques
are hindered by the relative intractability of closed string field theory
\cite{csft1,csft2,csft3,csft4,csft5,csft6}.  
The action itself is nonpolynomial, and closed string tachyons couple to the dilaton 
and the metric, making it necessary to carefully account for the backreaction of the tachyon 
condensate on the background.  Progress has been made in computing the bulk closed 
string tachyon potential perturbatively in the strength of the tachyon 
\cite{csftvac1,csftvac2,csftvac3,csftvac4,Moeller:2006cw,Moeller:2007mu} 
see also \cite{local9,Bergman:2004st} for analogous work
on localized tachyons), and some
evidence has emerged that a critical point of the potential may exist \cite{csftvac5,moeller}.

\heading{Worldsheet CFT}
A number of results have demonstrated 
that substantial progress can be made in understanding bulk closed string tachyon 
condensation directly as RG flow in the worldsheet CFT 
(e.g.,~\cite{Bergman:2005qf,paper4,Aharony:2006ra,paper2,paper1,paper0,paper3,paper5,paper6}).
Concrete conclusions can be reached by focusing on a class of exact solutions of the string theory 
in which quantum corrections are tightly constrained and calculable to all orders in 
perturbation theory.  
The simplest examples arise in bosonic string theory when the tachyon condenses 
with a purely exponential profile varying in a null direction:
\be
T \sim \exp{\b X^+} \ ,
\ee
where $\b$ is constant.
These theories are particularly
straightforward, since the tachyon vertex operator is nonsingular in the vicinity of 
itself, and the diagrammatic structure of the CFT indicates that the theory is exact at tree level
\cite{paper1,paper2}.  

To linearized order in the deformation, 
the tachyon couples to the $2D$ CFT as a potential, and the onset of null tachyon 
condensation can be modeled on the worldsheet as the nucleation of a bubble-like
region of nonzero tachyon.  Inside this region, string states see a potential wall that
rapidly increases into the future, and at late times the bubble itself expands outward 
from the nucleation point at the speed of light \cite{paper1}.  The resulting 
picture is an expanding region of tachyon condensate from which all string 
states are expelled.  Since no physical degrees of freedom persist inside the bubble at late
times, this configuration has been called the ``bubble of nothing,'' similar in spirit
to the Witten instanton solution described in \cite{wittenbubble}.
The expectation is that dynamical spacetime ceases to exist deep inside 
the bubble.  

A related process was found to drive dynamical dimensional reduction (or ``dimension quenching''),
wherein high potential walls from the tachyon condensate 
localize one or more (but not greater than $D-2$, where $D$ is the dimension of 
spacetime) spatial coordinates on the worldsheet
\cite{paper0,paper2}.
By again adopting a null tachyon profile, quantum corrections can be computed exactly at
finite loop order in perturbation theory.  In essence, the worldsheet fields that feel
the potential become infinitely massive and decouple from the theory.  Classically, this
amounts to a deficit in central charge contribution from the worldsheet degrees of freedom. 
This discrepancy is resolved at the quantum level by the simultaneous one-loop 
renormalization of the dilaton gradient and the string frame metric.  

Stated succinctly, both of these processes confirm qualitatively the expectation that
a natural consequence of bosonic closed string tachyon condensation is the spontaneous decay of 
spacetime.  On the worldsheet this is manifested as the creation of regions of ``nothing,''
where exponentially growing potential walls prevent the presence of string states,
and several examples of this process are now well understood in the 
language of worldsheet conformal field theory.

\heading{Effective actions}
If the intuition coming from the worldsheet CFT is correct,  
the two-dimensional theory should give rise to 
very interesting phenomena in spacetime (see, e.g.,~\cite{BergmanRazamat,Yangbigcrunch}).  
However, the spacetime dynamics in the presence of the tachyon 
configurations discussed above are not well understood.  
Such solutions fall into a class
of exponential tachyon profiles evolving in the background of a linear dilaton, and 
it is unclear whether the spacetime effective actions typically studied in the literature
consistently support these systems.

In this paper we aim to examine bulk closed string tachyon condensation directly as a dynamical 
cosmological process in spacetime.  
While this question is most easily and directly studied at the level of an effective action, 
it is difficult to asses the reliability of this 
approach without having a unique expression for the potential in which the fields of interest evolve.  
Furthermore, even if a potential is known, it is still unclear whether a low-derivative truncation, 
for example, is sufficient to capture the worldsheet physics of tachyonic fields.   

To address these questions, we adopt a simple strategy. 
Since it seems unlikely that a single two-derivative
effective action will consistently support all known tachyonic solutions of bosonic 
string theory, we aim to focus on a specific but nontrivial class of
solutions that contains controllable models of bulk tachyon condensation.  
Namely, the theories of interest are linearized deformations of the exact supercritical
$(D>26)$ linear dilaton background of bosonic string theory,
characterized by the insertion of purely exponential tachyon vertex operators.
We use these solutions
to guide the formulation of a consistent two-derivative effective action, 
including a specific form for the 
tachyon potential.  

The tachyon perturbations of interest are formulated at linearized order
in conformal perturbation theory.  
In other words, the tachyon profiles under consideration obey a linearized equation of 
motion.  In general, the solutions capture the dominant behavior of the string theory
when the system is perturbative in the strength of the tachyon.  
However, we show that one can rely on exact solutions of the 
worldsheet CFT to study the effects of higher-order corrections on the spacetime dynamics 
outside this perturbative regime.  
For the null tachyon system in 
particular, higher-order contributions to the action can be captured in the spacetime 
solution in a single undetermined function of the tachyon.  We demonstrate that 
when this function is chosen such that non-gravitational singularities are forbidden to 
appear in the action, and the gravity sector is constrained to be unitary, all possible 
curvature singularities are either removed or placed at $T=\infty$.

In the next section we review the tachyonic solutions of interest, and
demonstrate that a unique two-derivative effective action can be computed 
that supports the aforementioned class of solutions 
perturbatively in the strength of the tachyon.  In Section \ref{cosmo} we
study the process of 
tachyon condensation as cosmological evolution in an FRW target space.
Focusing on a timelike tachyon profile ($T \sim \exp{-\b^0 X^0}$),
we show that, modulo higher-order corrections, the evolution of the 
tachyon is realized in spacetime as a big crunch occurring in finite time, 
at arbitrarily weak string coupling.
(Numerical integration of the spacetime equations of motion for the dilaton, tachyon and 
FRW scale factor verifies that the effective action reproduces the timelike tachyon solution
quantitatively in the region of classical validity.)
We also analyze the exact null tachyon solution in the background of a timelike
linear dilaton rolling to weak coupling.  In this case, the worldsheet bubble of nothing
is realized in spacetime as a logarithmically expanding region in which the scale
factor collapses to a singularity.  In Section \ref{resolved}, we consider the null tachyon
solution on general grounds, and demonstrate that the naive gravitational singularity can be 
resolved classically.  The resulting system provides a very
simple mechanism by which the flow of cosmological time is either initiated or 
halted by string theory.  This suggests a class of toy models of the Big Bang, which 
can be studied at weak coupling.  We extend this analysis in Section \ref{solitons} 
to study solitonic configurations that localize the universe 
along a spatial direction.

\section{The effective action}
\label{effact}
Our initial goal is to derive an effective action that perturbatively supports a full class
of tachyonic solutions of bosonic string theory. 
The solutions of interest are linearized tachyonic deformations of the linear dilaton
CFT, characterized by the insertion of purely exponential tachyon vertex operators.  
The linear dilaton background is taken to be that of supercritical bosonic string theory defined 
in $D>26$ spacetime embedding dimensions, labeled by $X^\m,~\m \in 0,1,\ldots,D-1$.

In obtaining a consistent effective action, we will allow nontrivial
functions of the matter fields to appear multiplying the Einstein-Hilbert term.  It will
therefore be useful to introduce the following nomenclature when referring to different 
reference frames of the effective action:
\begin{itemize}
\item In {\bf sigma-model frame}, the metric $G^\s_{\m\n}$ is that which appears naturally in
	the $2D$ worldsheet CFT.  In all of the solutions of interest, the sigma-model
	metric will be that of flat, $D$-dimensional Minkowski space.  In this
	frame, the Einstein-Hilbert term does not appear canonically, though the dilaton
	dependence of the effective action appears as an overall factor of $\exp{-2\Phi}$.
\item {\bf String frame} will refer to the frame in which the Einstein-Hilbert term 
	in the effective action appears with just a factor of $\exp{-2\Phi}$ (while the 
	collective dilaton dependence of the action remains as an overall factor of $\exp{-2\Phi}$).
\item In {\bf Einstein frame}, the Einstein-Hilbert term appears canonically, and the 
	prefactor $\exp{-2\Phi}$ is removed by Weyl transformation.
\end{itemize}
When possible ambiguity arises, the frame will be specified by the sub- or superscript labels
$\s$, $S$, and $E$.  At the classical level, none of these frames is preferred in principle over 
the others (since there is no equivalence principle).
To analyze the cosmological aspects of the solutions at hand, however, it is easiest and most
intuitive to work either in string frame or Einstein frame.

When the sigma-model metric is flat $(G_{\mu\nu}^\s = \eta_{\m\n})$, 
the linear dilaton background alone comprises an exact solution with vanishing $B$-field, 
in which the worldsheet path integral is precisely Gaussian, and the (constant) dilaton gradient
$v_\m \equiv \pp_\m \Phi$ satisfies 
\be
v\cdot v = -\frac{D-26}{6 \apr} \ .
\label{vcondition}
\ee
In other words, the worldsheet beta functions for this background 
vanish to all orders in $\apr$:
\be
\beta^{G^\s} = \beta^\Phi = \beta^B = 0 \ .
\ee
The $2D$ energy-momentum tensor, 
in worldsheet lightcone coordinates $\s\uu\pm = - \s\uu 0 \pm \s\uu 1$, is
\be
\ct_{++} &=& - {1\over{\apr}} :\pp_{\s\uu +} X\uu\m \pp_{\s\uu +} X_\m :
+ \pp\sqd_{\s\uu +} (v_\m X\uu\m)\ ,
\nn\\ \nn\\
\ct_{--} &=& - {1\over{\apr}} :\pp_{\s\uu -} X\uu\m \pp_{\s\uu -} X_\m :
+ \pp\sqd_{\s\uu -} (v_\m X\uu\m)\ .
\ee

We now wish to introduce bulk tachyonic deformations of the linear dilaton CFT.
To linear order in the field strength, the deformation is implemented
by the insertion of a single tachyon vertex operator into all
correlation functions.  The perturbation is marginal as long as the matter 
part of the vertex operator is constrained to appear as a conformal primary
of weight $(1,1)$.   (In other words, the tachyon vertex operator can be 
made Weyl invariant, with the above energy-momentum tensor.)
This is achieved at linear order by satisfying the on-shell condition
\be
\pp^2 {T}(X) - 2\, v^\m\, \pp_\m {T}(X) + {4\over{\apr}} {T}(X) = 0\ .
\label{tachyonmarginality}
\ee

The strategy is to focus on the general class of solutions to this equation given by:
\be
T(X) = \mu \, \exp{ \b_\m X^\m} \ , 
\qquad
v_\mu = {\rm const.},
\qquad
G^\s_{\m\n}  =  \eta_{\m\n} \ ,
\label{solutions}
\ee
where $\mu$ is a free, real parameter (not to be confused with the spacetime index), 
and both $v_\m$ and $\b_\m$ are 
constant, $D$-dimensional vectors.  In general, the solutions in this
class are neither exact in the $\apr$ expansion, nor in conformal perturbation theory.
Even in the $\apr\to 0 $ limit, we expect that the all-orders dynamics of the effective theory
can become strongly corrected relative to the linearized approximation when the 
tachyon is of order one.

\subsection{Particular solutions}
Two particular solutions in the class described above will play an important role
in the analysis that follows.  The {\it timelike tachyon} is defined by the profile
\be
T(X^0) = \mu\, \exp{ - \b^0 \, X^0 } \ .  
\ee
We will study this solution in the background of a timelike dilaton
\be
\Phi = -v^0 \, X^0  \equiv -q\, X^0\ .
\label{timelike}
\ee
Since the timelike dilaton profile appears in several places, it is convenient to 
assign $v^0 \equiv q$.
The dilaton component of the worldsheet beta function equations demand that
\be
q = \pm \sqrt{\frac{D-26}{6\apr}} \ .
\label{branch}
\ee
By choosing the branch of this equation that renders $q$ positive definite
in $D>26$, the string coupling decreases toward the future.  
In turn, the tachyon marginality condition requires
\be
\b^0 = q \pm \sqrt{\frac{4}{\apr} + q^2} \ .
\ee
Arranging the tachyon to increase toward the future (i.e.~requiring $\b^0$ to be
negative definite), and substituting the solution for $q$ from above, one obtains
\be
\b^0 = \frac{  \sqrt{D-26} - \sqrt{D-2}  }{6\, \apr} \ .
\label{b0solution}
\ee
The parameter $\mu$ amounts to a shift of $X^0$, so we can study the condensation 
of the timelike tachyon at arbitrarily weak string coupling by taking
$\mu \ll 1$.   

\begin{figure}[htb]
\begin{center}
\includegraphics[width=2.8in,height=2.8in,angle=0]{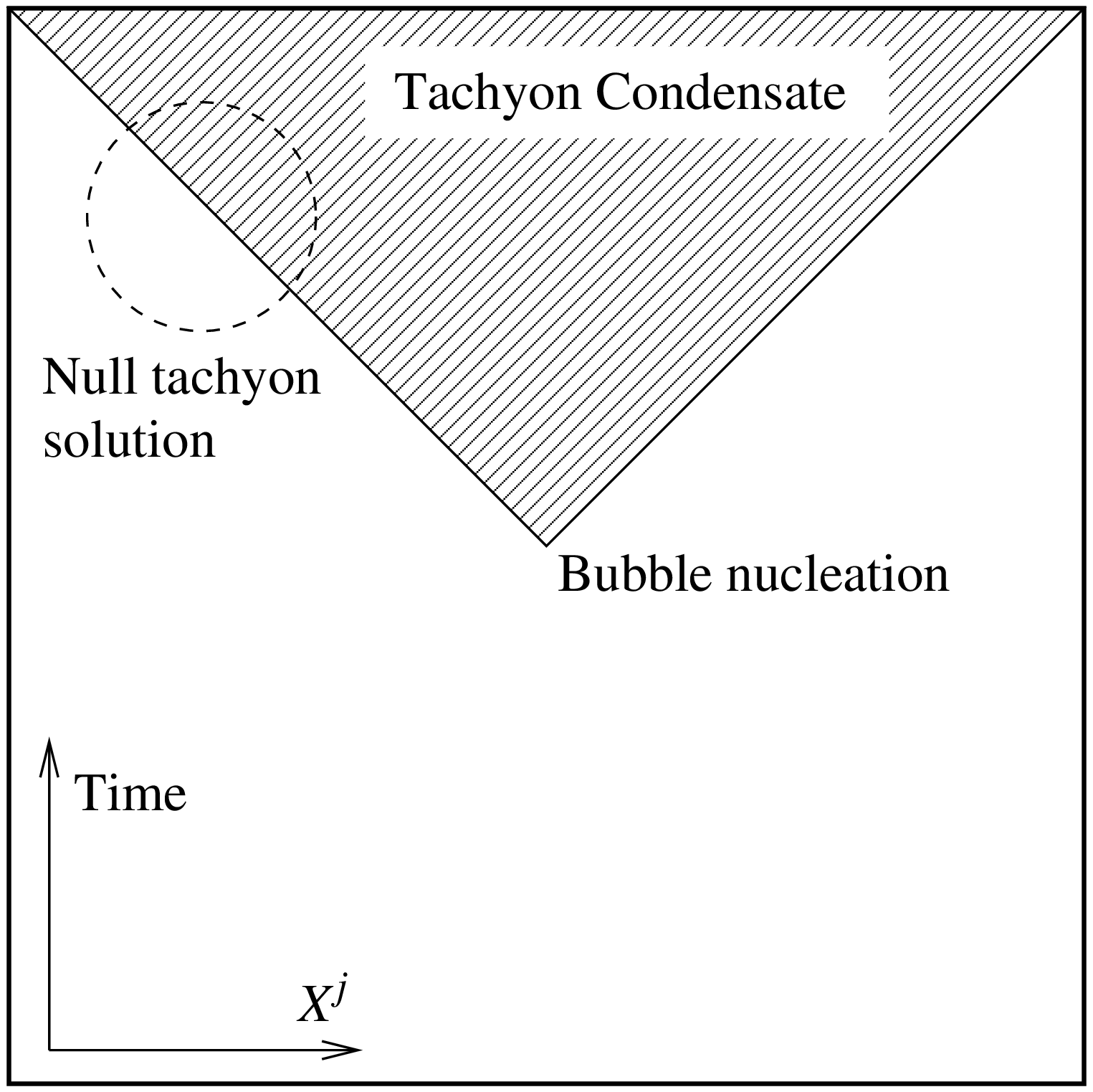}
\caption{Schematic diagram of the worldsheet bubble of nothing.  
The null tachyon solution can be thought
of as focusing on a region to the left of the origin of the $X^j$ coordinate.  
Physical degrees of freedom do not persist deep inside the
tachyon condensate.   }
\label{condensate}
\end{center}
\end{figure}

The second solution that will be examined below is the {\it null tachyon}, characterized by
the profile
\be
T(X^+) = \mu \, \exp{ \b_+ X^+ } = \mu\, \exp{ \frac{\b_+}{\sqrt{2}}(X^0 + X^j)}\ ,
\label{intronull}
\ee
where $X^j$ is an arbitrary spatial direction ($j\in 1,2,\ldots,D-1$).
In the presence of the timelike linear dilaton background, 
the tachyon equation of motion is satisfied when
\be
\b_+ = \frac{2\sqrt{2}}{q \apr} \ .
\ee
Once again we can choose the strength of the tachyon to increase into the future, in
the direction of weak string coupling, which amounts to selecting the $q > 0$ branch
of Eqn.~\rr{branch}.  
In this case, however, the tachyon is constant along lightfronts for which
$X^+ = {\rm const.}$  We can think of this solution as an approximate description of a bubble
of tachyon condensate, which nucleates on the worldsheet and expands outward at the 
speed of light (see Fig.~\ref{condensate}).  Since the tachyon vertex operator couples as
a potential in the worldsheet sigma model, the growth of the tachyon is manifested as the
appearance of a diagonal Liouville wall.  String states are prevented from passing deep into
the potential wall, so at late times the theory exhibits an expanding region that is 
completely devoid of physical degrees of freedom.  

\subsection{Higher-order corrections}
Since the goal of this study is to analyze closed string tachyon condensation
in regimes where the classical description is reliable, 
it is useful to briefly review the various sources
of higher-order corrections that can arise in the effective theory.  
First, higher genus corrections become strong when the string coupling $g_S$ is of order one.	
The effective string coupling is defined by $g_S = \exp{\Phi}$, so the classical limit corresponds 
to a dilaton expectation value of $\Phi = -\infty$.  All of the solutions of interest
exhibit a tachyon profile that increases in the direction of decreasing string coupling, and the
entire region of non-infinitesimal tachyon can be placed at arbitrarily small $g_S$. 
For the purposes of the present analysis, therefore, we will focus strictly on the 
weakly coupled regime.

Working to linearized order in conformal perturbation theory, the exact linear dilaton CFT
is deformed by the insertion of a single tachyon vertex 
operator into all correlation functions.  To this order, the conformal invariance of the 
tachyon deformation is encoded by the tachyon equation of motion 
\rr{tachyonmarginality}.  In general, the insertion of multiple vertex operators will 
lead to singularities, and conformal invariance at higher order will be restored 
by corrections to the linearized equation of motion.  At the level of the effective theory, 
conformal perturbation theory corresponds to an expansion of the action in nonlinearities of
fields.

Finally, the contribution of higher-dimension operators to the 
worldsheet beta equations translates in spacetime to the appearance of
higher-dimension operators in the effective action.  These operators are
suppressed by corresponding powers of $\apr$, and the classical limit
is reached by taking $\apr \to 0$, which is the limit of infinite string tension.

With respect to the above corrections, 
the null tachyon solution turns out to be remarkably simple at the level of the worldsheet CFT.
Consider the $2D$ worldsheet theory with 
general tachyon profiles of the form $T(X) = \m\, \exp{\b_\m X^\m}$.  For general
constant $\b^\m$, the insertion of two tachyon vertex operators will lead
to singularities when the positions of the operators become coincident.  
To be precise, singularities of normal-ordered operators arise in a free theory 
when propagators contract free fields in one operator with those of a second operator.
In the case of the tachyon vertex operators considered here, operators depend
only on $\b_\m X^\m$, so all field contractions, and all higher-order corrections to
the linearized tachyon equation of motion, will be proportional to $\b_\m \b^\m$.
For the null tachyon, however, $\b_\m \b^\m$ vanishes identically.

Furthermore, when the null tachyon solution is
expressed in lightcone coordinates, it is straightforward to demonstrate \cite{paper1,paper2}
that: 1) the propagator for the fields $X^\pm$ is oriented, directed from $X^+$
to $X^-$, and 2) all interaction vertices in the theory depend only on $X^+$.
As such, there are no Feynman diagrams beyond tree level, and hence no quantum loop corrections
whatsoever.\footnote{This is possible because the theory is not unitary prior to enforcing 
conformal gauge constraints.}  
Taken together, the above facts indicate that the linearized tree-level solution \rr{intronull}
is exactly conformally invariant.  
We will rely on this fact below to study the general spacetime dynamics of the null tachyon.

It turns out that, in large spacetime dimension, the timelike tachyon solution 
exhibits properties similar to the null tachyon.   
From Eqn.~\rr{b0solution}, one obtains
\be
\b_\m \b^\m = - \frac{4}{\apr^2\,D} + O(1/D^2) \qquad {\rm (timelike\ tachyon)}\ ,
\ee
which vanishes in the $D\to\infty$ limit. 
Furthermore, it is straightforward to show that worldsheet loops are also suppressed 
at large $D$ (see, e.g.,~\cite{paper0}).  
Near $D = \infty$ and $g_S \ll 1$, therefore, higher-order corrections to the timelike 
tachyon solution \rr{timelike} are strongly suppressed.

\subsection{String frame effective action}
The basic approach to constructing a spacetime effective action of the 
worldsheet CFT of string theory is to find an action whose equations of motion encode the
condition that all Weyl anomalies in the $2D$ theory vanish.  
Here we will restrict the analysis to an action 
containing terms with at most two spacetime derivatives; higher dimension 
terms are suppressed by higher powers of $\apr$.
Since our initial goal is to study the spacetime physics of the tachyonic solutions in
Eqn.~\rr{solutions}, which are tree-level solutions of the worldsheet beta function
equations, the hope is to capture the classical physical content of these solutions within 
the framework of a two-derivative effective action.

The simple class of tachyonic solutions described above \rr{solutions} is not
directly supported by the form of the spacetime effective action most often studied in the
literature.  The essential additional ingredient is that the Einstein-Hilbert term must 
appear with a nontrivial tachyon-dependent prefactor if the theory
is to simultaneously support the solutions of interest and 
admit a non-vanishing tachyon potential.  Without the latter, 
there would of course be no hope of reproducing tachyon scattering amplitudes, for example,
at the level of the effective action.

Our starting point will therefore be the most general two-derivative effective
action for the dilaton, metric and tachyon \cite{paper1}.
We begin in sigma-model frame with the following generic form:
\be
S = \frac{1}{2\kappa^2}\int d^D x \sqrt{-\det G^\s}\Bigl[
	{\cal F}_1 R^\s - {\cal F}_2 (\nabla \Phi)^2 - {\cal F}_3 (\nabla { T})^2
	- {\cal F}_4 - {\cal F}_5 \nabla { T} \cdot \nabla \Phi \Bigr]\ .
\label{stringframe}
\ee
The coupling $\kappa$ is related to the Newton constant by $G_N = \kappa^2 / 8\pi$.
Tree-level string amplitudes are defined such that the
dilaton dependence of the tree-level action appears as
an overall factor of $\exp{- 2\Phi}$, and
we have absorbed this prefactor into each of the functions
${\cal F}_ i$.  Apart from this factor, the ${\cal F}_ i$ are understood to be completely
arbitrary functions of the tachyon. 
For convenience, and to align conventions with the literature, we 
encode the explicit tachyon dependence of these functions via the following definitions:
\be
{\cal F}_1  \equiv  e^{-2\Phi}\, f_1({ T})\ , 
&\qquad &
{\cal F}_2  \equiv  -4\, e^{-2\Phi}\, f_2({ T})  \ , 
\xxx
{\cal F}_3  \equiv  e^{-2\Phi}\, f_3({ T})  \ , 
& \qquad & 
{\cal F}_4  \equiv  2\, e^{-2\Phi}\, V_\s({ T})  \ , 
\xxx
{\cal F}_5  \equiv  e^{-2\Phi}\, f_5({ T})  \ .
&\qquad &
\label{bigf}
\ee
The functions $f_i(T)$ on the right-hand side of Eqn.~\rr{bigf} depend only on the tachyon,
including the potential $V_\s(T)$.

In terms of these functions, the Einstein equation appears as
\be
0 & = & \left(\nabla^\mu \nabla^\nu - G_\s^{\mu\nu} \nabla^2 
	+ \frac{1}{2} G_\s^{\mu\nu} G_\s^{\rho\sigma} R_{\rho\sigma}
	- R^{\mu\nu}\right) {\cal F}_1
	- \frac{1}{2}G_\s^{\mu\nu}{\cal F}_4
	+ {\cal F}_2\pp^\mu \Phi \pp^\nu \Phi
\nn\\
&&	- \frac{1}{2}G_\s^{\mu\nu} {\cal F}_2 (\pp \Phi)^2
	+ {\cal F}_3 \pp^\mu { T} \pp^\nu { T}
	- \frac{1}{2} G_\s^{\mu\nu} {\cal F}_3 (\pp { T})^2
	- \frac{1}{2}{\cal F}_5 G_\s^{\mu\nu} \pp_\rho { T} \pp^\rho \Phi
\nn\\
&&	+ \frac{1}{2}{\cal F}_5 \pp^\mu { T} \pp^\nu \Phi
	+\frac{1}{2} {\cal F}_5 \pp^\nu { T} \pp^\mu \Phi\ ,
\label{einsteinequation}
\ee
where $\nabla^\mu$ is the usual covariant derivative.
The dilaton and tachyon equations of motion read, 
respectively:
\be
0 & = & -2 R f_1 + 8 f_2 (\pp \Phi)^2
	- 8 f_2' \pp { T} \cdot \pp \Phi
	- 8 f_2 \nabla^2 \Phi + (2 f_3 + f_5') (\pp { T})^2
	+ f_5 \nabla^2 { T} + 4\, {V_\s}\ ,
\nn\\
&&
\label{dilatoneom}
\\
0 & = & f_1'\, R + (4 f_2' - 2 f_5)(\pp \Phi)^2 + f_3' (\pp { T})^2
	- 4 f_3 \pp \Phi \cdot \pp {T} - 2\, {V_\s}'
	+ f_5 \nabla^2 \Phi + 2 f_3 \nabla^2 { T}\ ,
\nn\\
&&
\label{tachyoneom}
\ee
where $f_i'(T) \equiv \pp_T f_i(T)$.

We now require that the effective action in Eqn.~\rr{stringframe}
support the class of tachyonic solutions of interest, given in Eqn.~\rr{solutions} 
above.  (See, e.g.,~\cite{Kutasov:2003er} for a similar technique, employed in open string theory.)
Projecting onto these solutions, the Einstein equation \rr{einsteinequation} 
stipulates the following constraint:
\be
0 &=&  \eta_{\m\n} \biggl[
	2\, V_\s + 4\, (2\, f_1 - f_2) v\cdot v 
	+ T\, \left[
	f_5\, v\cdot \b + 2\, f_1' (\b\cdot \b - 4\, v \cdot \b)
	+ (2\, f_1'' + f_3) \, \b \cdot \b \,T \right] \
	\biggr]
\xxx
&&	+  8\, (f_2 - f_1)\, v_\m v_\n
	+ T\, \left[
	(v_\m \b_\n + \b_\m v_\n )(4\, f_1'-f_5)
	- 2\, \b_\m \b_\n \, \left( f_1' + (f_1'' + f_3) \, T\right)
	\right] \ . 
\label{cond1}
\ee
The tachyon equation of motion becomes
\be
0 = (2\, f_2' - f_5)\, v\cdot v - \pp_T V_\s - 2\,f_3 \,T \,v\cdot \b 
	+ \b\cdot \b\, T \left(f_3 + \frac{1}{2} f_3' \,T\right) \ ,
\label{cond2}
\ee
while the dilaton equation of motion gives
\be
0 = 8 \,f_2 \,v\cdot v + 4 V_\s - 8 \,f_2' \,T\, v\cdot \b + f_5\, T\,  \b\cdot\b 
	+ (2\,f_3 + f_5')\, T^2\, \b\cdot \b \ .
\label{cond3}
\ee
The above equations (\ref{cond1},~\ref{cond2},~\ref{cond3}) 
are uniquely satisfied in terms of the
lone function $f_1(T)$ by the following solution:
\be
f_2(T) & = & f_1(T) \ ,
\xxx
f_3(T) & = & -f_1''(T) - \frac{f_1'(T)}{T} \ ,
\xxx
f_5(T) & = & 4\, f_1'(T) \ ,
\xxx
V_\s(T) & = & -\frac{1}{2}\left(
	4\, f_1(T) v\cdot v + T\, f_1'(T)\,(\b\cdot\b - 4\, v\cdot \b)
	+ T^2\,f_1''(T)\,\b\cdot\b \right) \ . 
\label{prepot}
\ee

From the solution for $f_3(T)$, we see that for the tachyon kinetic term to be finite 
at $T=0$, we must have that 
\be
f_1'(T) \Bigr|_{T=0} = 0 \ .
\ee
In addition, we also require the conformal invariance of the exponential 
tachyon perturbation to linearized order in conformal perturbation theory 
(i.e.,~that it satisfies the on-shell condition in
Eqn.~\rr{tachyonmarginality} above): 
\be
\b\cdot\b - 2\, v\cdot\b  + \frac{4}{\apr} = 0 \ .
\ee
Imposing this, along with the condition that the dilaton component of the beta function equations
vanish \rr{vcondition}, we can eliminate $v^\m$ from the potential \rr{prepot}. 
One recovers the following form:
\be
V_\s(T) = \frac{1}{2} \biggl[ 
	4 \,f_1(T)\, \left(\frac{D-26}{6\apr}\right)
	+ T\,f_1'(T)\,\left( \b\cdot\b + \frac{8}{\apr}\right)
	- T^2\, f_1''(T)\,\b\cdot\b 
	\biggr] \ .
\label{betadependence}
\ee

At this stage, we could demand that the action be independent of the form of any particular solution.
The condition that the potential be completely independent of $\b^\m$  
translates to a condition on the function $f_1(T)$ of the form
\be
\pp_{\b\uu\m} V(T) = {\b_\m} \left( T\, f_1'(T) - T^2\,f_1''(T) \right) \equiv 0\ .
\ee
So for nonzero tachyon and nonzero $\b_\m$, we recover the following form for the
function $f_ 1(T)$:
\be
f_ 1(T) = c_0 + c_1\, T^2 \ ,
\ee
where $c_1$ and $c_2$ are free constant parameters.

There is another route to this result.  
In general, we should impose that {\it any} solution to the
equations of motion must preserve conformal invariance to leading order in
perturbation theory.  For a general background, this means that the system should satisfy
\be
\nabla^2 T - 2\,\pp_\m\Phi \pp^\m T + \frac{4}{\apr} T = 0 \ .
\label{generalmarginality}
\ee
In other words, for solutions of the effective theory, 
the conditions of conformal invariance imposed 
by the $2D$ worldsheet theory should emerge as a prediction, rather than an input, of the 
effective action.  In essence, we demand that the spacetime effective action correctly reproduce 
the known worldsheet tachyon scattering amplitudes to leading order
in conformal perturbation theory and $\apr$.  To be sure, higher-order corrections
can become important for completely general backgrounds.  For the purposes of this
calculation we can consider imposing \rr{generalmarginality} in the presence of 
small deviations from solutions that are known to exist in a perturbative regime.

For general $T$ and $\Phi$, the dilaton and tachyon 
equations of motion take the form (dropping the explicit $T$ dependence of $f_1(T)$):
\be
0 & = & f_1' \nabla^2 T - 2\, f_1' \pp\Phi\cdot \pp T
	+ \frac{f_1''}{2} \pp T \cdot \pp T
	- \frac{f_1}{2} \left(4 \nabla^2 \Phi - 4 \pp\Phi \cdot \pp\Phi + R\right)
	- \frac{f_1'}{2 T} \pp T\cdot \pp T + V_\s(T) \ ,
\xxx
0 & = & \left[ f_1' - T(f_1'' + f_1''' T)\right] \pp T \cdot \pp T
	+ T \Bigl[
		f_1' T \left(4 \nabla^2 \Phi - 4 \pp\Phi \cdot \pp\Phi + R \right)
\xxx
&&	
	+ 2(f_1' + f_1'' T)( 2 \pp\Phi\cdot  \pp T -   \nabla^2 T)
	- 2 T\, \pp_T V_\s (T) 
	\Bigr]\ .	
\label{f1EOM}
\ee
Note that the Einstein equation simplifies significantly when the dilaton equation is enforced
(specifically, terms proportional to $G^\s_{\m\n}$ drop out entirely):
\be
2 f_1\, \nabla_\m \nabla_\n \Phi 
	- f_1'\, \nabla_\m \nabla_\n T 
	+ \frac{f_1'}{T}\, \pp_\m T \pp_\n T
	+ f_1\, R_{\m\n} = 0 \ .
\label{EINSTEIN}
\ee
Using the trace of this equation, the dilaton equation of motion also simplifies
\be
4\, f_1\, \pp\Phi\cdot \pp\Phi - 4\,f_1'\, \pp\Phi\cdot \pp T
+ f_1\, R + \left(f_1'' + \frac{f_1'}{T}\right)\, \pp T\cdot \pp T + 2\, V_\s = 0 \ .
\ee

Employing the dilaton equation of motion \rr{f1EOM} to eliminate $\nabla^2 \Phi$, and imposing the 
condition \rr{generalmarginality}, the tachyon equation of motion
yields the following condition on $f_1(T)$:
\be
T^2\,f_1\,f_1''' - \left( f_1 - T\, f_1' \right)\left(f_1' - T\, f_1'' \right) = 0\ .
\label{f1constraint}
\ee
At the level of the effective theory, conformal perturbation theory corresponds to an
expansion in the strength of the tachyon field.  Since the condition is that conformal
invariance is imposed at leading order, we can solve this equation perturbatively in $T$:
\be
f_1(T) = \sum_{n=0}^\infty c_n \, T^n \ .
\ee
The function $f_1(T)$ appears as a coefficient of the Einstein-Hilbert term in the
effective action
\be
S_{\rm EH} \sim \int d^D X \, e^{-2\Phi}\, f_1(T)\, R^\s \ ,
\ee
so we require $c_0 = 1$ for the theory to properly reduce to the unperturbed linear
dilaton background when the tachyon vanishes.  Furthermore, the leading order contribution from 
\rr{f1constraint} establishes that $c_1 = 0$.  It is then clear, working order-by-order, that
Eqn.~\rr{f1constraint} is satisfied perturbatively to all orders in $T$ for any $c_2$, and 
\be
c_k = 0 \ , \qquad \forall\, k > 2 \ .
\ee
Furthermore, for the tachyon potential 
to be tachyonic, the constant $c_2$ must be negative definite.   It turns out
that the magnitude of $c_2$ drops out of the entire action under a constant rescaling
of the tachyon.  Setting $c_2 = -1$, we recover the solution
\be
f_1(T) = 1 - T^2 \ .
\label{f1solution}
\ee

We pause to emphasize an important aspect of this result.  As noted above, the solutions under 
consideration \rr{solutions} are, in general, not exact. 
For the tachyon perturbation to remain conformally invariant to
higher orders in conformal perturbation theory, the linearized tachyon equation
\rr{generalmarginality} will inevitably acquire nonlinear corrections.  
These corrections will ultimately alter the condition on the 
function $f_1(T)$ in Eqn.~\rr{f1constraint}.  
We should therefore not regard the solution in Eqn.~\rr{f1solution}
as exact to all orders in the strength of the tachyon:  
\be
f_1(T) = 1 - T^2 + O(T^3) \ .
\label{f1exp}
\ee
The resulting tachyon potential in sigma-model frame takes the following form:
\be
V_\s(T) = \frac{1}{3\apr}\left( (D-26) - (D-2)\, T^2 \right) + O(T^3)\ .
\ee
Happily, the potential is now completely independent of the vector $\b^\m$ to the order of interest, 
rendering the effective theory independent of any particular solution.

For the moment, we wish to study the leading-order dynamics of this theory
in certain special cases, 
and we will momentarily drop all reference to higher-order 
corrections in the small-$T$ expansion (though we will return to this issue in Section \ref{resolved}).
At this stage, the spacetime action takes the form
\be
S &=& \frac{1}{2\kappa^2} \int d^D X\,\sqrt{-\det G^\s} \,e^{-2\Phi}\, \biggl[
	(1-T^2)\, R^\s
	+ 4\, (1-T^2)\, \pp_\m\Phi \pp\uu\m\Phi
	+ 8\, T\, \pp_\m T \pp\uu\m\Phi
\xxx
&&	-4\, \pp_\m T \pp\uu\m T
	+\frac{2}{3\apr}\left(26-D+ (D-2)T^2\right) 
	\biggr]\ .
\ee

To canonicalize the gravity sector, we can invoke a spacetime Weyl transformation
\be
G^S_{\m\n} = e^{2\,\o(T) } G^\s_{\m\n} \ ,
\label{weyl1}
\ee
where $G^S_{\m\n}$ is the string frame metric, and 
\be
\o(T) =  \frac{\log (1-T^2) }{D-2} \ .
\label{weyl2}
\ee
In the next section, however, we will move completely over to Einstein frame.  
We will therefore combine the above field redefinition with an additional Weyl 
transformation that renders the Einstein-frame action in canonical form.

\section{Spacetime cosmology}
\label{cosmo}
In this section we focus on the dynamics of the timelike and null tachyon solutions, as encoded 
by the spacetime effective action.  As noted above, 
we expect the semiclassical analysis to be reliable when the strength of the tachyon is small
compared to one.   The goal is to study the general behavior of these 
solutions in regions of classical validity, but we also wish to establish that, in the absence of
higher-order corrections, singularities eventually arise as a consequence of tachyon condensation.

\subsection{Translation to Einstein frame}
Because the Einstein-Hilbert term in the sigma-model action appears with the prefactor 
$f_1(T)$, the Weyl rescaling that renders this term canonical in Einstein frame depends 
on both the dilaton and the tachyon: 
\be
G^E_{\m\n} \equiv e^{2\o(\Phi,T) } G^\s_{\m\n} \ .
\label{conformaltform}
\ee
Under this field redefinition, the Ricci term in Einstein frame takes the form
\be
S_{\rm EH} \sim \int d^D X\, \sqrt{-\det G^E}  
	\, f_1(T)\, \exp{ (2-D)\, \o(\Phi,T) - 2\Phi } R^E \ .
\ee
We therefore find the following expression for $\o(\Phi,T)$:
\be
\o(\Phi,T) = \frac{-2\,\Phi + \log (1-T^2) }{D-2} \ .
\ee
At this stage, it is easy to see that as the tachyon magnitude evolves 
from some small initial value at early times to 
$T = \pm 1$, the Einstein metric inevitably encounters a big crunch in finite time.
However, if the tachyon evolves slowly from zero, the metric can reach this singularity 
deep within a region where the string theory is weakly coupled.  
When $T^2$ increases beyond unity, the Einstein term in sigma-model frame 
changes sign, and the conformal transformation \rr{conformaltform} becomes 
imaginary.   This can be interpreted as a signal that the gravitational 
theory becomes nonunitary for $T^2 > 1$.   

It is interesting that the theory encounters a singularity at the point
where we expect the dynamics to acquire strong corrections relative to 
the linearized approximation \rr{solutions}.  Returning to this issue in Section
\ref{resolved}, we will consider the ability of higher-order corrections to 
resolve the singularity.  If such corrections contribute to
higher-order terms in $f_1(T)$ \rr{f1exp}, the tachyon dependence in the
metric will obviously change.  For now, however, we aim to study
the action as it stands, leaving open the question of higher-order corrections.  
Our goals in this section are 1) to establish that this
singular region in fact arises in the classical analysis, 
and 2) to test whether the effective action reliably reproduces the classical 
solutions of interest \rr{solutions} away from singular points.  

It is straightforward to carry out the transformation in Eqn.~\rr{conformaltform} 
on the remaining terms in the action.  To canonicalize the dilaton kinetic term, 
we perform an additional rescaling:
\be
\Phi = \frac{1}{2}\sqrt{D-2}\, \phi \ .
\ee
We recover the following two-derivative effective action in Einstein frame:
\be
S &=& \frac{1}{2\kappa^2}\int d^D X \, \sqrt{-\det G^E}\, \biggl[
	R^E 
	-\pp_\m \phi \pp\uu\m \phi
	-\frac{4}{\sqrt{D-2}} \frac{T}{(1-T^2)} \pp_\m\phi \pp\uu\m T
\xxx
&&	-\frac{4(D-2+T^2)}{(D-2)(1-T^2)^2} \pp_\m T\pp\uu\m T
	-\frac{2}{3\apr} (1-T^2)^{-\frac{D}{D-2}} 
		\left( D-26-(D-2)T^2 \right) e^{\frac{2\phi}{\sqrt{D-2}}}
	\biggr] \ .
\nn\\
&&
\label{einsteinframeaction}
\ee
One can see that the singularity in the metric at $T=\pm 1$ is also translated to the
kinetic terms.

\begin{figure}[htb]
\begin{center}
\includegraphics[width=2.8in,height=2.5in,angle=0]{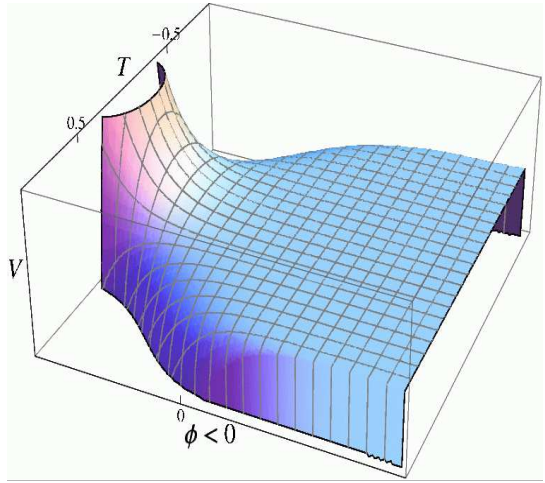}
\caption{The dilaton-tachyon potential $V_E(\phi,T)$ in Einstein frame 
(depicted at $D=30,~\apr=1$).  
When the tachyon passes the critical value $\pm T^*$, the potential decreases as
a function of $\phi $ in the direction of strong coupling. }
\label{potential}
\end{center}
\end{figure}

We have thus found a specific form of the dilaton-tachyon potential
$V_E(\Phi,T)$ in Einstein frame:
\be
V_E(\Phi,T) = \frac{1}{3\apr} (1-T^2)^{-\frac{D}{D-2}} 
		\left( D-26-(D-2)T^2 \right) e^{\frac{2\phi}{\sqrt{D-2}}} \ .
\label{perturbativeeinsteinpotential}
\ee
The potential is depicted for $D=30$ in Fig.~\ref{potential}.
From this picture it is easy to understand the generic behavior of the system evolving
from a state with zero tachyon.  At the outset, the theory evolves toward weak coupling
as the dilaton rolls down its potential toward decreasing negative values (to the right in 
Fig.~\ref{potential}).  Small fluctuations of the tachyon eventually cause it to roll toward
positive or negative magnitude, reaching $T = \pm 1$ asymptotically. 

At zero tachyon, the potential
increases exponentially in the direction of increasing positive $\phi $
(to the left in Fig.~\ref{potential}, which is the direction of strong string coupling).  
There is a critical magnitude of the tachyon 
\be
T^* = \sqrt{1 + \frac{24}{2-D}} \ ,
\ee
at which the potential vanishes identically.  When the tachyon is above this magnitude,
the potential is negative and decreases exponentially as a function of 
$\phi$ in the direction of strong coupling.  At weak coupling the potential is
essentially flat, falling off steeply at $T=\pm 1$.

\subsection{Timelike tachyon}
We now want to focus in particular on the timelike tachyon:
\be
T = \mu\, \exp{ - \b^0 \, t_\s } \ ,  
\qquad
\Phi = \frac{1}{2}\sqrt{D-2}\, \phi = -q \, t_\s  \ ,
\qquad
G^\s_{\m\n} = \eta_{\m\n} \ .
\label{timelike2}
\ee
We have labeled the sigma-model time coordinate as
$X^0_\s \equiv t_\s$.  Again, by choosing the dilaton gradient  $q$ to be positive definite
in $D>26$, the string coupling decreases toward the future.

To study the cosmology associated with these solutions directly, we will proceed
by moving to a coordinate system in which the Einstein metric is of FRW form:
\be
ds_E^2 = -dt_E^2 + a(t_E)^2 d \vec X_E^2 \ .
\label{FRW1}
\ee
We wish to study the timelike tachyon on shell, and
the conformal rescaling in Eqn.~\rr{conformaltform} provides a precise translation 
between FRW coordinates $X^\m_E$ and sigma model coordinates $X_\s^\m$:
\be
dt_E^2 =  e^{2\o(\Phi,T)} dt_\s^2 \ ,
\qquad
a(t_E)^2 d\vec X_E^2 =  e^{2\o(\Phi,T)} d\vec X_\s^2\ .
\ee
(For simplicity we are keeping the time dependence of the dilaton and tachyon implicit.)

We can keep the translation among spatial coordinates trivial
(i.e.,~$X^i_E = X^i_\s,~i\in 1,2,\ldots,D-1$)
by assigning 
\be
a(t_E)  =  \exp{\o(\Phi,T)} = \exp{ \frac{-2 \Phi + \log (1-T^2)}{D-2}  }\ .
\label{scalefactor}
\ee
This leaves the translation between timelike coordinates explicit.
When the tachyon is zero, we recover
\be
t_E(t_\s) = \frac{D-2}{2q}\, \exp{ \frac{2q}{D-2} t_\s } + {\rm const.},
\qquad (T = 0) \ .
\ee
After the tachyon acquires a vev, the translation for general dilaton
and tachyon profiles takes the form 
\be
t_E(t_\s) =  \int_1^{t_\s} d\xi \, e^{ -2\Phi(\xi) / (D-2)} 
	\left(1 - T(\xi)^2\right)^{1/(D-2)}   + {\rm const.}
\ee
For the solution at hand \rr{timelike2}, we obtain
\be
t_E(t_\s) &=& \frac{D-2}{2(q-\b^0)} \exp{\frac{2 q t_\s}{D-2}}
	(1-\L(t_\s))^{\frac{1}{2-D}} (1 - \L(t_\s)^{-1})^{\frac{1}{D-2}}
\xxx
&&
	\times \, {}_2 F_1 \left(
	\frac{1}{2-D} , \frac{q - \b^0}{\b^0(D-2)}
	, \frac{q + \b^0(D-3)}{\b^0(D-2)}
	, \L(t_\s) \right) + {\rm const.},
\ee
where $\L(t_\s)$ is defined by
\be
\L(t_\s) \equiv \frac{1}{\m^2}\,{e^{2\,\b^0\, t_\s}} \ .
\ee
${}_2 F_1 (a,b,c,z)$ is the hypergeometric function, which exhibits a branch cut 
in the complex $z$ plane along the real $z$ axis from $z = 1$ to $\infty$.

It turns out that we can re-express the solution $t_E(t_\s)$ as
\be
t_E(t_\s)  =  {\m^{  \frac{ 2 q }{\b^0 (D-2) } }} \,
	\frac{ e^{-\frac{i\pi}{D-2}} }  {2\b^0} \,
	B_{\L(t_\s)} \left( \frac{ q - \b^0 }{\b^0(D-2)} ,  \frac{D-1}{D-2}\right)
	+ {\rm const.},
\label{betasolution}
\ee
where ${B}_z(a,b)$ is the incomplete Euler beta function
$B_z(a,b) = \int_0^z d\xi  (1-\xi)^{b-1}\,\xi^{a-1} $. 
The beta function ${B}_z(a,b)$ also exhibits a branch cut, though in this
case it runs along the negative real 
$z$ axis from $z=-\infty$ to $z=0$.  As it stands, $t_E(t_\s)$ (modulo the integration constant) 
contributes a constant  imaginary piece when $t_\s$ lies in the region 
prior to the final singularity.  In the analysis that follows it will be understood
that this contribution is subtracted by absorbing it into the integration constant.

The timelike tachyon solution can thus be expressed as a function of the time coordinate
in Einstein frame by inverting Eqn.~\rr{betasolution} to generate
$t_\s$ as a function of $t_E$.  There is not a convenient closed-form expression
for $t_\s(t_E)$, however.  Keeping this relationship implicit, one obtains
\be
T(t_E) = \m\, e^{-\b^0\, t_\s(t_E)  } \ .
\label{analytictime1}
\ee 
Likewise, the dilaton evolves according to
\be
\Phi(t_E) = \frac{1}{2}\sqrt{D-2}\,\phi = - \sqrt{ \frac{D-26}{6\apr} } \, t_\s (t_E)  \ .
\label{analytictime2}
\ee
Substituting into the general form for the scale factor in 
Eqn.~\rr{scalefactor}, we find
\be 
a(t_E) = \exp{ \frac{  2\, q\, t_\s(t_E) + 
	\log ( 1- \m^2 \, e^{-2 \,\b^0\, t_\s(t_E)  }  ) }{D-2}  } \ .
\label{analytictime3}
\ee
By construction, this is an exact (albeit particular) classical solution to the 
equations of motion of the spacetime effective action in Einstein frame.

\begin{figure}[htb]
\begin{center}
\subfigure[Scale factor]{
\includegraphics[width=2.5in,height=1.8in,angle=0]{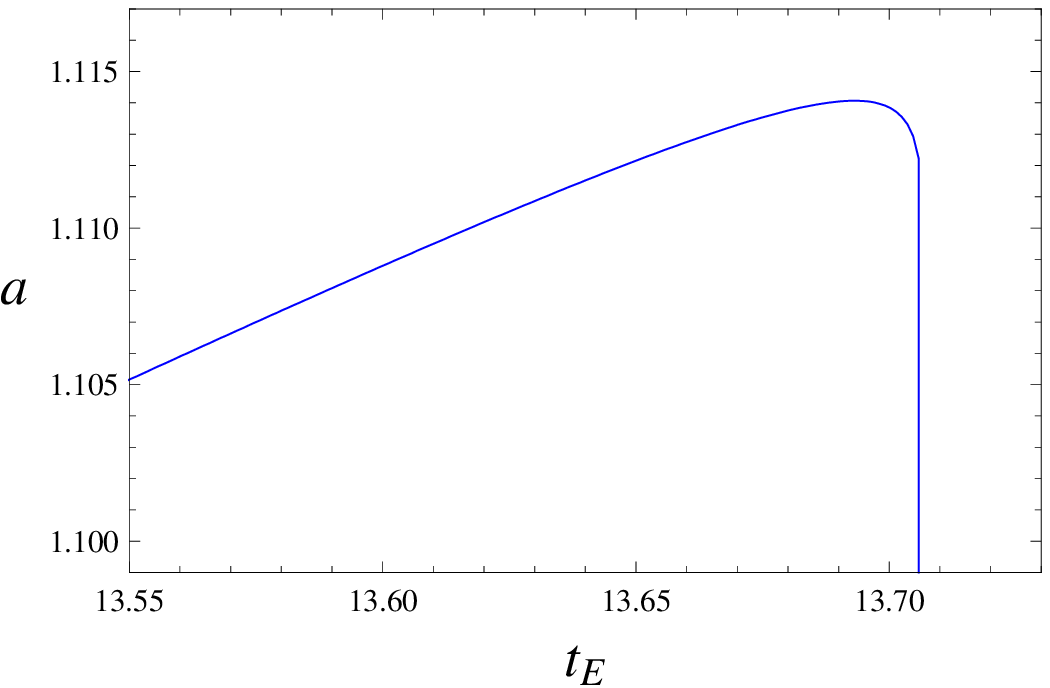}}
\hspace{8mm}
\subfigure[String coupling]{
\includegraphics[width=2.5in,height=1.8in,angle=0]{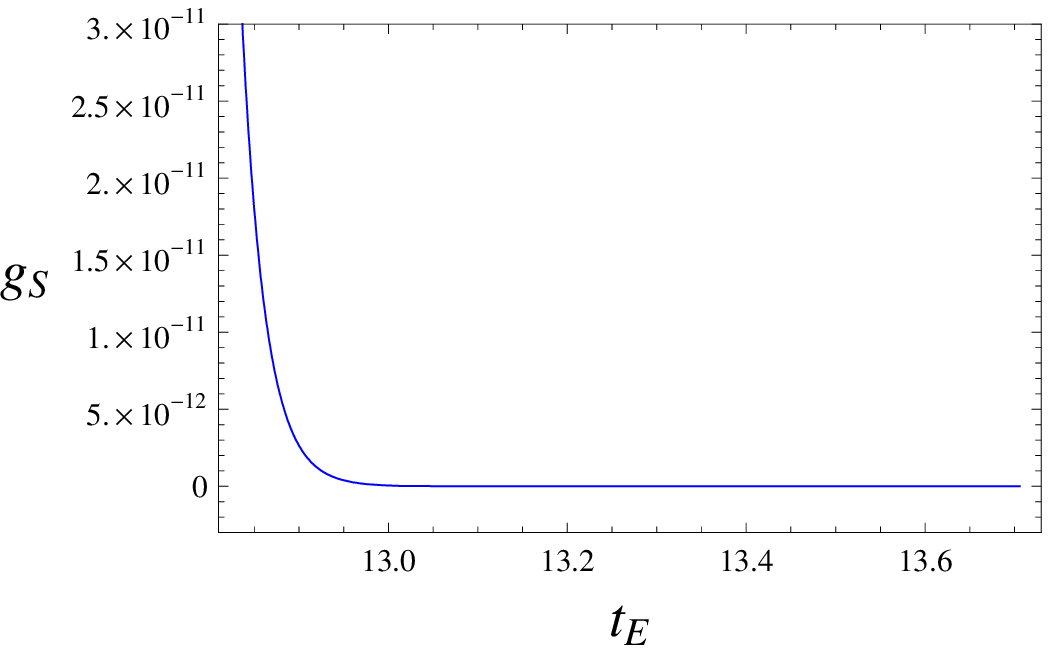}}
\caption{The evolution of the timelike tachyon ($\m = 1/2,~D=1000$).
The scale factor (panel (a)) evolves linearly as a function of time in Einstein frame
while the tachyon is small (i.e.,~it evolves at a critical equation of state).  As the
magnitude of the tachyon increases, the scale factor enters a phase of accelerated
contraction, and eventually reaches a singularity.  The string coupling (panel (b)) 
decreases throughout this process.}
\label{examplesolution}
\end{center}
\end{figure}

It is straightforward to plot the behavior of the scale factor numerically.  This is done
in Fig.~\ref{examplesolution},
along with the corresponding evolution of the string coupling.
(Here, and in the analysis that follows, $\apr$ can be set to any convenient value without
affecting the qualitative behavior of the solutions.)
When the tachyon is small, the scale factor evolves linearly as a function of $t_E$.
As the tachyon evolves away from zero, the universe enters a phase of accelerated contraction.
Eventually, as the tachyon strength approaches one, the scale factor collapses to zero.
The region of non-negligible tachyon condensate can exist entirely 
within a region of weak string coupling $g_S \approx 0$.  

In sigma-model coordinates, the tachyon magnitude reaches one at
\be
t_\s^{\rm crunch} = \frac{1}{\b^0} \log \mu = \frac{6\, \apr\,\log \mu}{  \sqrt{D-26} - \sqrt{D-2}  } \ .
\ee
In Einstein frame, this translates to the statement that the scale factor formally
collapses to a singularity at the time
\be
t_E^{\rm crunch} =  \frac{\m^{  \frac{ 2 q }{\b^0 (D-2) } }}{2\b^0}  
	\cos \left(\frac{\pi}{D-2}\right)  
	B \left( \frac{ q - \b^0 }{\b^0(D-2)} ,  \frac{D-1}{D-2}\right) \ ,
\ee
where the expression on the right-hand side is given in terms of the complete Euler beta function 
\be
B(a,b) = \G(a)\G(b)/\G(a+b) \ .
\ee

\begin{figure}[htb]
\begin{center}
\subfigure[Scale factor]{
\includegraphics[width=2.5in,height=1.8in,angle=0]{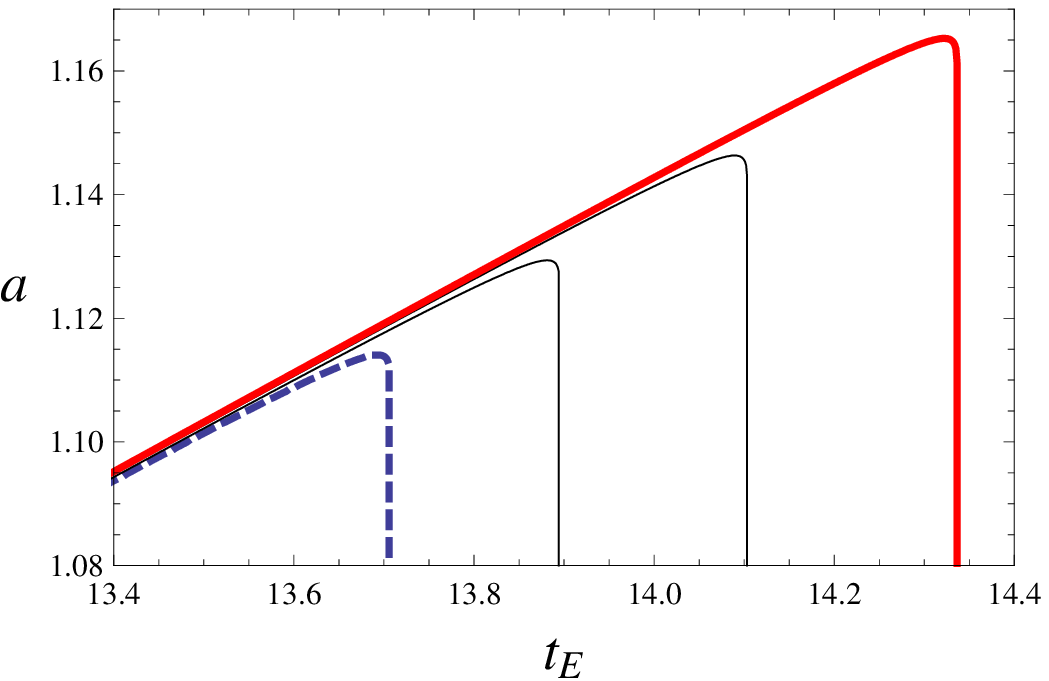}}
\hspace{8mm}
\subfigure[Tachyon magnitude]{
\includegraphics[width=2.5in,height=1.8in,angle=0]{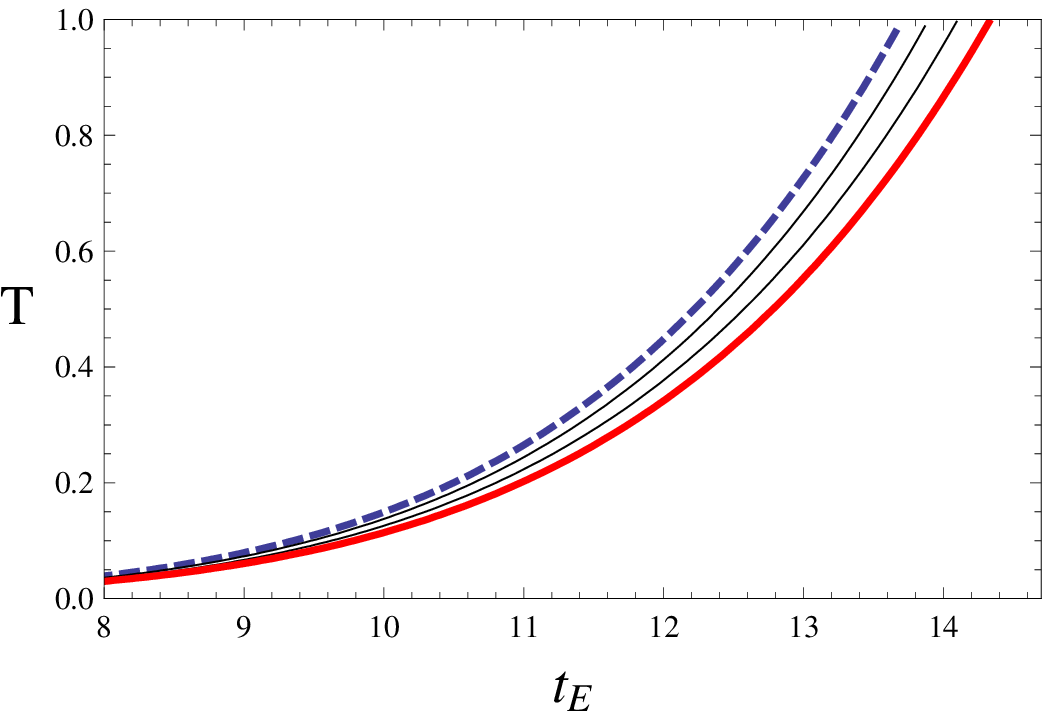}}
\subfigure[String coupling]{
\includegraphics[width=2.5in,height=1.8in,angle=0]{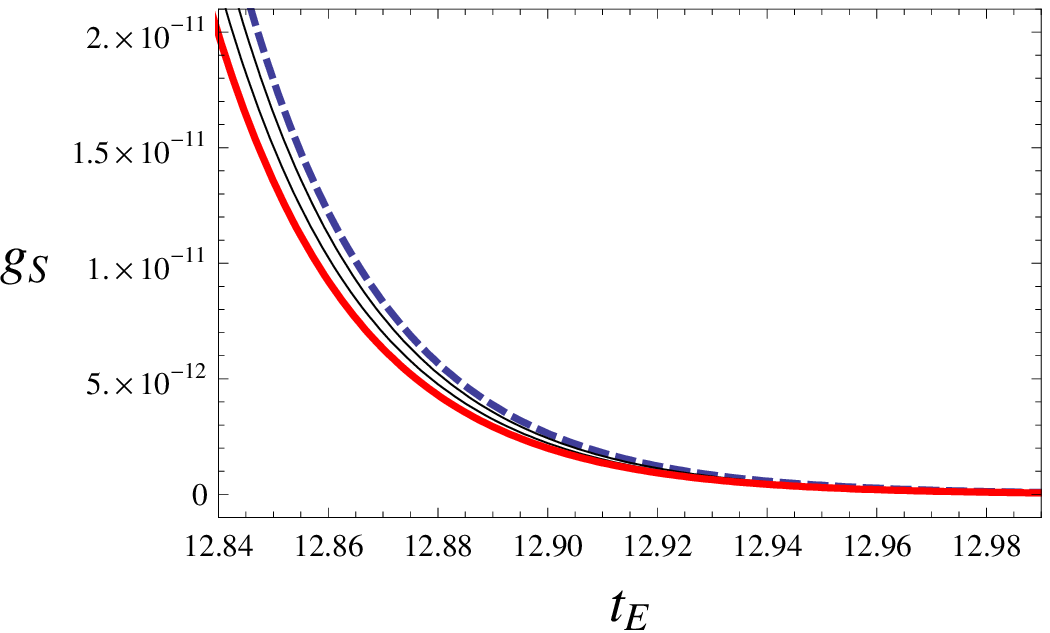}}
\caption{The timelike tachyon system from $\m = .38$ (red-solid) 
to $\mu= .5$ (blue-dashed), with $D=1000$.  The intermediate solid curves are generated at $\mu=.42$ and $\mu=.46$.
The effect of decreasing $\mu$ is to place the final curvature singularity 
deeper in the direction of decreasing string coupling.  For each value of $\mu$
the scale factor (panel (a)) collapses in a big crunch precisely when the tachyon 
(panel (b)) reaches one. This process can occur throughout a region in which the string coupling
(panel (c)) is small.   }
\label{analyticsolutions}
\end{center}
\end{figure}

For fixed spacetime dimension, we have a one-parameter family of solutions of the timelike 
tachyon system, parameterized by the constant $\mu$.  In Fig.~\ref{analyticsolutions} 
we plot four such solutions for $\m = \{0.38,~0.42,~0.46,~0.5\}$ in fixed dimension. One can
see explicitly that by decreasing $\mu$, the tachyon reaches $T\approx 1$ more gradually, 
placing the final singularity of the scale factor farther out in the direction of decreasing 
string coupling.

\begin{figure}[htb]
\begin{center}
\includegraphics[width=2.5in,height=1.8in,angle=0]{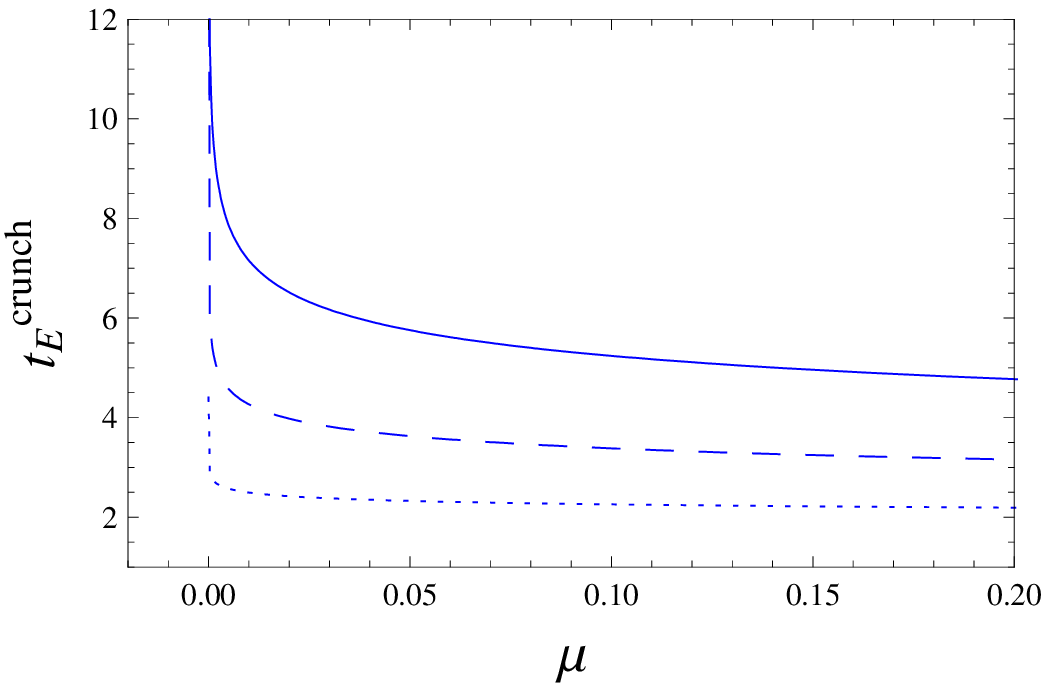}
\caption{The time at which the universe reaches a crunch in Einstein frame is prolonged
by decreasing $\mu$, or by increasing $D$.  The curves above are depicted with $D = 30$ (dotted curve),
$D=50$ (dashed curve) and $D=100$ (solid).}
\label{muplot}
\end{center}
\end{figure}

The relationship between 
$\mu$ and $t_E^{\rm crunch}$ is depicted for various $D$ in Fig.~\ref{muplot}.
From this plot we can also see that in higher spacetime dimension the crunch is placed at weaker
coupling for fixed $\mu$.  
To this is added the additional effect that, for a fixed time $t_E$, increasing the
dimension $D$ alone reduces the string coupling (at large $D$, the dilaton
gradient scales as $q \approx - \sqrt{ D }$).

It is straightforward to compute the spacetime equations of motion in Einstein frame.
If, for example, we are interested in studying the 
behavior of the timelike tachyon in the background of a timelike linear dilaton,
we can proceed with the ansatz that both the tachyon and dilaton fields depend 
only on the Einstein-frame time coordinate $t_E$.  
Furthermore, we can work in the coordinate system given in
Eqn.~\rr{FRW1}, in which the
metric $G_{\m\n}^E$ takes the form of a spatially flat ($k=0$) FRW metric:
\be
ds^2_E = -dt_E^2 + a(t_E)^2 d\vec X^2 \ .
\ee
With these assumptions, variation of the dilaton in the action yields
\be
0 &=& 	\ddot \phi + (D-1) H \dot \phi +  \pp_\phi V(\phi,T)
\xxx
&&		+ \frac{2}{\sqrt{D-2}} \frac{1}{ (1-T^2) } 
		\biggl[
		(D-1) H\, T\, \dot T 
		+ T\, \ddot T + 
		\frac{1+T^2}{1-T^2}\, \dot T^2
		\biggr] \ ,
\label{timelikeeom1}
\ee
where $H$ is the Hubble parameter $H \equiv {\dot a}/{a}$.
Varying the tachyon gives
\be
0 & = & \frac{1}{2}\sqrt{D-2}\left(
		\ddot \phi + (D-1)\,H\, \dot\phi \right)\,T
	+ \frac{ D-2+T^2}{(1-T^2)} \left( \ddot T + (D-1)\, H \,\dot T\right)
\xxx
&&	+ \frac{2D-3+T^2}{(1-T^2)^2}\, T\,\dot T^2
	+ \frac{1}{4}(D-2)(1-T^2)\, \pp_T V(\phi,T) \ .
\label{timelikeeom2}
\ee
The pressure $p$ and energy density $\rho$ of the background take the form
\be
p & = & \frac{1}{2\kappa^2}\biggl[
	\dot\phi^2 + \frac{4}{\sqrt{D-2}} \frac{T}{(1-T^2)} \,\dot\phi\,\dot T
	+ \frac{4(D-2+T^2)}{(D-2)(1-T^2)^2} \,\dot T^2 - 2 V(\phi,T)
	\biggr] \ ,
\xxx
\rho & = &  p + \frac{2}{\kappa^2} V(\phi,T)\ .
\ee
The Einstein equations then reduce to 
\be
\frac{\ddot a}{a} = -\kappa^2 \rho\, \frac{ (D-3) + (D-1) w}{(D-2)(D-1)} \ ,
\label{timelikeeom3}
\ee
where $w$ is the usual equation of state $w \equiv p/\rho$.  We note that the critical 
equation of state, defining the boundary between an accelerating and decelerating 
cosmology, is \cite{paper1}
\be
w_{\rm crit} = -\frac{D-3}{D-1}\ .
\label{wcrit}
\ee
In other words, the scale factor accelerates as a function of $t_E$ if the 
equation of state lies in the range $-1 \leq w < w_{\rm crit}$.

At this point we can integrate Eqns.~(\ref{timelikeeom1},~\ref{timelikeeom2},~\ref{timelikeeom3})
numerically, given a set of initial conditions.  As noted above, 
solutions in the worldsheet theory fall into a family parameterized by $\mu$
(for fixed $D$ and $\apr$).  While this parameter is absent from the point of view of
the effective theory alone, the information contained in $\mu$ can be translated to
the effective dynamics in the form of integration constants.

\begin{figure}[htb]
\begin{center}
\includegraphics[width=2.5in,height=1.8in,angle=0]{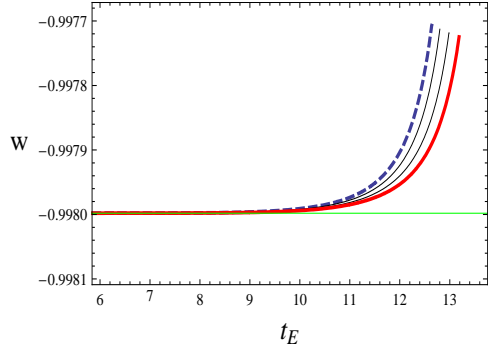}
\caption{The evolution of the  equation of state for
the timelike tachyon solutions under consideration (at $D=1000$).
The individual curves correspond to values of $\mu$ from $\mu=.38$ (red-solid) to
$\m=.5$ (blue-dashed).  The intermediate solid curves are generated at
$\m=.42$ and  $\mu=.46$.  The critical equation of state, given by 
$w_{\rm crit} = -(D-3)/(D-1)$, is depicted by the horizontal green curve. }
\label{hubbleandEOS}
\end{center}
\end{figure}

For the particular solutions under consideration, it is instructive to 
plot the equation of state as the system evolves toward the final singularity.  
This is displayed in Fig.~\ref{hubbleandEOS}.  It is easy to see that as the tachyon
remains small, each system evolves at the critical equation of state
\rr{wcrit} (marked by the green horizontal line).  As the tachyon evolves, each system acquires 
an equation of state lying above the critical value, indicating a phase of decelerating
scale factor.

\subsection{The null tachyon}
We now turn to the null tachyon solution in the same setting.  Our goal for the moment
is to characterize the singular region exhibited by the null tachyon that 
arises in the absence of higher-order corrections to the action.  (We will consider 
higher-order effects in the next section.)
In sigma-model coordinates, the null tachyon solution is specified by
\be
&&\kern-20pt
	T(t_\s,X^1_\s)  = \mu\, \exp{ \frac{\b_+}{\sqrt{2}}\, (t_\s + X^1_\s)} 
	\equiv \mu\, \exp{\b X^+}\ ,
\xxx 
&&\kern-20pt
\Phi(t_\s) = - q\, t_\s \ , 
\qquad 
G^S_{\m\n} = \eta_{\m\n}\ ,
\label{nullsolution}
\ee
where $X^1_\s$ is a transverse embedding coordinate in sigma-model frame.
To keep the notation concise, we have relabeled the constant lightcone vector as
$\b_+ \equiv \b$.  Consistency of the string theory requires
\be
q = \pm \sqrt{\frac{D-26}{6\apr}} \ ,
\qquad 
q\, \b = \frac{2\sqrt{2}}{\apr} \ .
\ee
Again, choosing the positive branch of the dilaton gradient yields a system in which the
dilaton rolls to weak coupling in the future while the tachyon grows exponentially
at fixed $X^1_\s$.

In Einstein frame, we will adopt a coordinate system in which the metric is again 
of FRW form:
\be
ds^2_E = -dt_E^2 + a(t_E, X^1_E)^2 \, dX^i_E dX^i_E \ ; 
\qquad i \in 1,2,\ldots,D-1 \ .
\ee
The scale factor $a(t_E, X^1_E)$ now depends both on the timelike direction
$t_E$ and the single spatial coordinate $X_E^1$.  
As before, we can choose the mapping between Einstein-frame and sigma-model coordinates
to be trivial for the spatial directions $X^i$: 
\be
X^i_E  =  X^i_\s \, \equiv \, X^i \ , \qquad i \in 1,2,\ldots,D-1 \ .
\ee
These relations imply the following dependence of the scale factor on 
the dilaton and tachyon:
\be
a(t_E,X^1) = \exp{ \frac{-2 \Phi + \log (1-T^2)}{D-2}  }\ ,
\ee
where $T$ is now a function of both $t_\s$ and $X^1$ \rr{nullsolution}. 
The relation
\be
dt_E^2 = e^{2\o (\Phi,T)} dt_\s^2 
\ee
gives the following mapping for the Einstein 
coordinate $t_E$:
\be
t_E(t_\s,X^1) =  
	-\frac{\mu^{-\frac{2\sqrt{2} q}{\b (D-2)}} }{\sqrt{2}\b}\, 
	\exp{\frac{i \pi  +  2\, q\, X^1 }{ 2-D } }
	\, B_{\L(t_\s, X^1)} \left(
		\frac{q\sqrt{2} + \b}{\b (2-D)} , \frac{D-1}{D-2} \right)
	+ {\rm const.} \ ,
\label{nulltimeeqn}
\ee
where
\be
\L(t_\s,X^1) \equiv \frac{1}{\m^2}\exp{ -\sqrt{2}\b (t_\s + X^1) } \ .
\ee
As with the timelike tachyon, we need to absorb a constant imaginary contribution 
into the integration constant on the right-hand side of Eqn.~\rr{nulltimeeqn}.

At this point we can see that the theory reaches a curvature singularity at 
\be
t_\s^{\rm crunch} = -X^1 - \frac{\sqrt{2}}{\b} \log \m \ .
\ee
In FRW coordinates, this equates to 
\be
t_E^{\rm crunch} =  
	-\frac{\mu^{-\frac{2\sqrt{2} q}{\b (D-2)}} }{\sqrt{2}\b}\, 
	\exp{\frac{ 2\, q\, X^1 }{ 2-D } } 
	\cos\left( \frac{\pi}{2-D} \right)
	\, B \left(
		\frac{q\sqrt{2} + \b}{\b (2-D)} , \frac{D-1}{D-2} \right) \ .
\ee
To an observer at fixed negative $X^1$, the singular region
appears to approach from the positive $X^1$ direction at a speed given by
\be
v_{\rm bubble} = (D-2) \sqrt{ \frac{3\, \apr}{2(D-26)}}\, \frac{1}{t_E} \ .
\ee

\begin{figure}[htb]
\begin{center}
\includegraphics[width=2.5in,height=2.0in,angle=0]{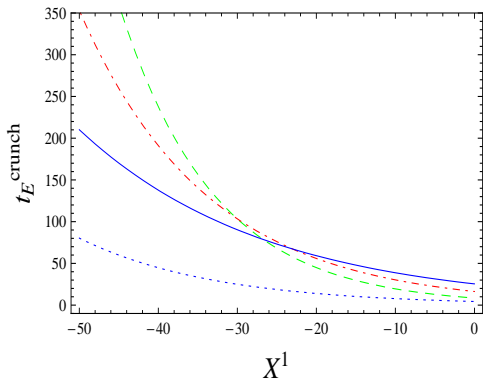}
\caption{The time at which the universe reaches the final crunch in the presence of the null tachyon,
as a function of $X^1$.  The curves shown above are depicted for $\mu = 0.5$,
$D=30$ (dotted, blue), $D=50$ (dashed, green), $D=150$ (dot-dashed, red) and $D=350$
(solid, blue). }
\label{nullcrunch}
\end{center}
\end{figure}

This picture is intuitively consistent with what we know from the $2D$ CFT.  
On the worldsheet, the growth of the tachyon appears as a potential
wall that increases exponentially with $X^+$ (in sigma-model coordinates),
preventing the presence of string states deep inside the region of tachyon
condensate.  To a rough approximation, we would expect spacetime to become 
nondynamical in a region that grows outward from the origin of the $X^1$ 
coordinate.  The nontrivial translation between sigma-model and Einstein-frame
coordinates modifies the picture somewhat, but the general outcome is as expected.
In Fig.~\ref{nullcrunch} we plot $t_E^{\rm crunch}$ as a function of $X^1$
for various $D$.   In spacetime, the bubble of nothing indeed emerges as a surface
of zero metric that expands outward in the direction of negative $X^1$.

\section{Resolved singularities}
\label{resolved}
The action under consideration thus far was derived under the constraint that it support the complete
class of tachyonic solutions defined in Eqn.~\rr{solutions}.  Terms in the effective action suppressed by
higher powers of $\apr$ have been systematically dropped, and the form of $f_1(T)$ was derived
at linearized order (in conformal perturbation theory).  In this section, we wish to consider the 
possible effects of higher-order corrections.  

\subsection{The null tachyon}
By focusing strictly on the null tachyon solution in the background of a
timelike linear dilaton rolling to weak coupling \rr{nullsolution}, we can study 
various higher-order corrections in isolation.
As noted above, the tree-level 
worldsheet solution is well-defined and exactly conformally invariant to all orders in 
perturbation theory, and nonperturbatively in $\apr$.  Furthermore, the effects of finite 
string coupling can be made arbitrarily small 
by placing the strongly coupled region deep in the past.

Since the null tachyon is exact in $\apr$, including higher-dimension operators in the
effective action consistently cannot lead to corrections to the spacetime equations of motion that are not 
automatically satisfied by the tree-level solution.  Similarly, while corrections associated with 
conformal perturbation theory will 
inevitably alter the tachyon marginality condition \rr{tachyonmarginality},  
such corrections should be satisfied trivially by the null tachyon solution.  However, the
constraint equation \rr{f1constraint} for $f_1(T)$  was derived for the non-exact tachyon 
profiles (plus fluctuations) appearing in Eqn.~\rr{solutions}.\footnote{Recall that the null 
tachyon is supported as a classical solution for {\it any} $f_1(T)$.}  
Corrections to the marginality condition will
ultimately contribute nontrivial corrections to Eqn.~\rr{f1constraint}.   
The conclusion is that the only higher-order effects that can appear directly in the 
spacetime null tachyon system arise as corrections to the 
function $f_1(T)$.\footnote{To be sure, analogous statements do not hold for non-exact solutions.}  
This raises the possibility that the curvature singularity that naively appears in the above 
analysis is resolved by higher-order effects.


\subsection{Generalized constraints on $f_1(T)$}
Without directly computing higher-order corrections to the effective action (a problem
that lies beyond the scope of this study),
we would like to understand on general grounds the terms that can arise as
corrections to $f_1(T)$, subject to the condition that the theory always support the null 
tachyon as an exact solution. 

Consider the effective action for general $f_1(T)$.  As it stands, 
this action is consistent 
with the condition that it support only the null tachyon as an exact solution.
Because of the null symmetry, the $\b$-dependence in the potential 
(see, e.g.,~Eqn.~\rr{betadependence}) is automatically absent:
\be
S & = & \frac{1}{2\kappa^2}\int d^D X \sqrt{-\det G^S} 
	\, e^{-2 \Phi} \biggl[
	R^S 
	- \frac{4}{{D-2}}  \left( \pp_\m\Phi \pp^\m \Phi
	+ \frac{f_1'}{f_1} \pp_\m\Phi \pp\uu\m T  \right)
\xxx
&&
	+ \frac{1}{f_1}\left[
		f_1'' + f_1'\left( 
		\frac{1}{T} 
		- \frac{(D-1)}{(D-2)}\frac{f_1'}{f_1}
		\right) 
		\right] \pp_\m T \, \pp\uu\m T
\xxx
&&
	- \frac{1}{3\apr} 
	f_1^{-\frac{D}{D-2}} \left( 24\, T\, f_1' + 2(D-26)\, f_1 \right)
	\biggr] \ .
\ee
Here the action appears in string frame, with the Einstein-Hilbert term 
expressed in canonical form.  For general $f_1(T)$, 
the string-frame metric is related to the sigma-model 
metric by
\be
G^S_{\m\n} = e^{2\o(T)} G^\s_{\m\n} 
\qquad \o(T) = \frac{\log f_1(T)}{D-2} \ .
\label{weylstringframe}
\ee
Of course, we can also move to Einstein frame using the Weyl transformation
\be
G^E_{\m\n} = e^{2\o(\Phi,T)} G^\s_{\m\n} \ ,
\qquad
\o(\Phi,T) = \frac{-2\,\Phi + \log f_1(T)}{D-2} \ .
\ee
For finite dilaton, singularities of the metric arise 
whenever $f_1(T)$ vanishes.\footnote{In sigma model frame, this corresponds to a
vanishing Einstein term in the action.}

We now want to study constraints on higher-order corrections to $f_1(T)$
under the following general conditions:
\begin{enumerate}
\item The theory encodes the standard, non-tachyonic linear dilaton background in the 
      limit $T=0$;
\item The gravity sector remains unitary for all values of $T$;
\item Non-metric prefactors of the matter sector kinetic terms remain finite for all values of $T$;
\item Expanding in the strength of the tachyon, $f_1(T)$ is defined to 
	quadratic order by $f_1(T) \approx 1 - T^2$. 
\end{enumerate}
The last condition guarantees that the potential in a canonical gravity frame is tachyonic.
If there is any hope that the effective action can reliably reproduce the physics
of the null tachyon
for all finite values of $T$, each of these conditions must be met.  
For the moment we leave open the possibility 
that the metric encounters a singularity as a consequence of tachyon condensation.

As noted in Section \ref{effact}, the first condition imposes
\be
f_1(0) = 1\ .
\ee
Furthermore, the condition that the gravity sector remain unitary can be satisfied 
by demanding that $f_1(T)$ be nonnegative for all $T$.  This is intuitive from the
perspective of the action in sigma-model frame, where the Ricci term appears
multiplied by $f_1(T)$.  
In addition, we see that for the tachyon kinetic term to remain finite at $T=0$, 
\be
f_1'(T) \Bigr|_{T=0} = 0 \ .
\ee
This indicates that $T=0$ is a critical point of $f_1(T)$.

The tachyon kinetic term (as well as the $\pp_\m T \pp^\m \Phi$ term) 
can also become singular at points where 
$f_1(T)$ vanishes.  The relevant factors are $f_1''(T) / f_1(T)$ and $f_1'(T)/f_1(T)$,
so we demand that $f_1'(T)$ and $f_1''(T)$ vanish whenever
$f_1(T) = 0$.  This implies that points where $f_1(T)$ vanishes must either be inflection
points, or points where $f_1(T)$ vanishes identically over a finite 
region.\footnote{An 
example of a $C^\infty$ function that is nontrivial in some region but vanishing in another
is $f_1(T) = \exp{-1/(a^2 - x^2)}\, \Theta(a^2-x^2)$, where $\Theta(x)$ is the 
step function.}  Since $f_1(T)$ is everywhere nonnegative, however, there can be no inflection points
coinciding with points where $f_1(T)$ vanishes.  Furthermore, if $f_1(T)$ vanishes 
identically over a finite region, the {\it entire} action becomes identically zero in this
region.  If we hope to reliably encode the dynamics of the complete string theory
for all values of the tachyon, we are forced to reject this scenario. 
We conclude that, under the second and third constraints above, $f_1(T)$ can never vanish
at finite values of $T$.  At finite $T$, therefore, the null tachyon
avoids {\it all} cosmological singularities.

\subsection{A candidate effective action}
We would now like to understand the relation between the above constraints on $f_1(T)$
and the perturbative form that was computed in Section \ref{effact}.  The constraint
imposed there was that the effective action should support all linearized tachyonic 
perturbations to the linear dilaton CFT of the form
\be
T = \mu\, \exp{\b_\m X^\m} \ ,
\ee
subject to the condition that the tachyon profile satisfies the (linearized) tachyon
equation of motion in Eqn.~\rr{tachyonmarginality}.  For solutions other than the null tachyon, 
and in the absence of an additional regulator (like large $D$, for example),
corrections to the linearized marginality condition can generate higher-order corrections to
the general constraint equation for $f_1(T)$ \rr{f1constraint}.  
As noted above, we should therefore regard the solution 
for $f_1(T)$ used in Section \ref{effact} as an approximation valid at small $T$:
\be
f_1(T) \approx 1- T^2 \ , \qquad T\ll 1 \ .
\ee
A natural additional constraint on the general function $f_1(T)$
is that it reproduce the perturbative expansion around small $T$ (i.e.,~condition (4) above).
This guarantees that the effective action supports the full class of solutions
in Eqn.~\rr{solutions} in regions where the solutions themselves are not strongly 
corrected by higher-order effects.  
A simple exact form for $f_1(T)$ that meets all of the above 
criteria and reproduces the known solution at small $T$ is
\be
f_1(T) = \exp{-T^2} = 1 - T^2 + O(T^4) \ .
\label{fullf1solution}
\ee

The resulting action in string frame takes the form (using the rescaled dilaton)
\be
S &=& \frac{1}{2\kappa^2} \int d^D X \sqrt{-\det G^S} \, e^{-\sqrt{D-2}\, \phi} \,
	\Bigl[
	R^S - \pp_\m \phi \pp^\m \phi
	-\frac{4}{\sqrt{D-2}} \,T\, \pp_\m \phi \pp^\m T
\xxx
&&
\kern-20pt
	-\frac{4}{D-2} \,(T^2 + D-2)\,	\pp_\m T \pp^\m T
	-\frac{2}{3\apr}\, \exp{\frac{2\,T^2}{D-2}} \, (D-26 - 24\,T^2)
	\Bigr] \ .
\label{neweffact}
\ee
As desired, the potential 
\be
V_S(T) = \frac{1}{3\apr} \, \exp{\frac{2\,T^2}{D-2}} \, (D-26 - 24\,T^2)
\ee
is tachyonic.  In this case, however, it is well-defined at $T=1$.  
It can easily be verified that this potential agrees with the tachyon potential 
computed above in Eqn.~\rr{perturbativeeinsteinpotential}, up to and including cubic order in an 
expansion around small $T$.

Of course, there are other possible
completions of $f_1(T)$ that meet all of the conditions described above.
Another example is
\be
f_1(T) = \frac{1}{\cosh (\sqrt{2}\,T)} = 1 - T^2 + O(T^4) \ ,
\ee
which also resolves the curvature singularity in the null tachyon solution
at finite $T$.  The string frame action takes the form
\be
S &=& \frac{1}{2\kappa^2} \int d^D X \sqrt{-\det G^S} \, e^{-\sqrt{D-2}\, \phi} \,
	\Bigl[
	R^S - \pp_\m \phi \pp^\m \phi
	+ \frac{2\sqrt{2}}{\sqrt{D-2}}\,\tanh (\sqrt{2}\,T)\,\pp_\m\phi \pp^\m T
\xxx
&&	-\left(
		\frac{2}{D-2} 
		+ \frac{2(D-3)}{(D-2)}\, {\rm sech}^2 (\sqrt{2}\,T)
		+ \frac{\sqrt{2}}{T}\, \tanh (\sqrt{2}\,T)
	\right)\,\pp_\m T \pp^\m T
\xxx
&&	-\frac{2}{3\apr}\,{\rm sech}^{-\frac{2}{D-2}} (\sqrt{2}\,T)
	\left( D-26 - 12\sqrt{2}\,T\,\tanh (\sqrt{2}\,T) \right)
	\Bigr] \ .
\ee

The qualitative properties of the null tachyon solution
are equivalent for both forms of $f_1(T)$ given above.  
For the purposes of the present study, therefore, we will
employ the exponential form in Eqn.~\rr{fullf1solution} 
in the analysis that follows.  While it would be interesting to 
study a wider class of solutions, we emphasize that the resolution of 
the crunch is universal to all $f_1(T)$ satisfying the above conditions.

\subsection{Resolution of the singularity}
On general grounds, we expect that the dynamics of the effective action in Eqn.~\rr{neweffact}
resolve the cosmological singularity that appears naively in the null tachyon solution.  
We now examine in detail how the singularity is avoided. 

The string frame metric is given in Eqn.~\rr{weylstringframe} above. 
We again choose the Weyl transformation to act trivially on the transverse 
spatial coordinate $X^1_\s$, so it is convenient to drop the subscript label.
The string frame time coordinate thus evolves as a function of $t_\s$ and $X^1$ according to
\be
t_S(t_\s,X^1) = \frac{1}{\sqrt{2}\b} \,
	{\rm Ei} \left( -\frac{\m^2}{D-2} e^{\sqrt{2}\,\b\, (t_\s + X^1)} \right)
	+ {\rm const.}\ ,
\label{resolvedstringtime}
\ee
where ${\rm Ei}(x)$ is the exponential integral function 
${\rm Ei(x)} = -\int_{-x}^\infty\, (e^{-\xi} / \xi)\, d\xi$
(Ei$(x)$ exhibits a branch cut in the complex $x$ plane running from $x=-\infty$ to $x=0$,
though for real $x$ we take the principal value of the integral).
For convenience we set the constant term to zero.

The (string-frame) FRW scale factor now takes the form
\be
a_S(t_\s,X^1) = \exp{ \frac{\log f_1(T)} {D-2} } = \exp{ -\frac{T(t_\s,X^1)^2}{D-2} } \ .
\ee
For the null tachyon, this becomes
\be
a_S(t_\s,X^1) = \exp{ -\frac{\m^2}{D-2} e^{\sqrt{2}\,\b\,(t_\s + X^1)} } \ .
\ee
Expressed in terms of sigma-model coordinates, the outcome is clear.  
For fixed $X^1$, the evolution of $t_\s$ drives an accelerated contraction of the 
scale factor (in the positive branch of \rr{branch}, $\b>0$).  However, 
$a_S(t_\s,X^1)$ can never reach a true singularity in finite $t_\s$, as the singular point
has been moved to $t_\s = \infty$.  

\begin{figure}[htb]
\begin{center}
\includegraphics[width=2.5in,height=1.6in,angle=0]{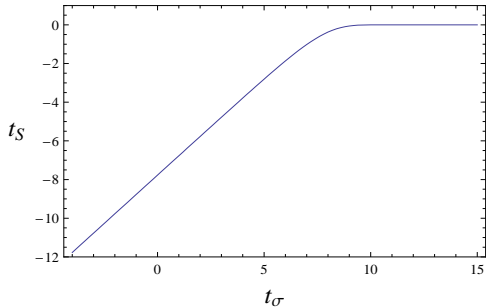}
\caption{The ``flow of time'' in string frame, as a function of $t_\s$.
Deep inside the tachyon condensate, the spacetime dynamics become infinitely redshifted
relative to the sigma model.   (The plot depicted is generated at 
$D=1000,~\m=0.5,~X^1=0$.)}
\label{stringtime}
\end{center}
\end{figure}

Interestingly, the dependence on $X^1$ drops out when the scale factor is expressed 
as a function of string-frame coordinates.  To see this, first note that Eqn.~\rr{resolvedstringtime}
can be inverted to give the sigma-model time coordinate as a function of $t_S$ and $X^1$:
\be
t_\s(t_S,X^1) = -X^1 
	+ \frac{1}{\sqrt{2}\,\b} \log \left( -\frac{D-2}{\m^2}\, {\rm Ei}^{-1} (\sqrt{2}\,\b\,t_S)
	\right) \ .
\ee
In string frame, the scale factor thus takes the explicit form
\be
a_S(t_S) = \exp{ {\rm Ei}^{-1} (\sqrt{2}\,\b\, t_S) } \ .
\ee
Naively, the system reaches a curvature singularity at $t_S = 0$.
Relative to the sigma-model frame, however, the physics in string frame is dramatically 
redshifted as the scale factor collapses.  We can see this directly by plotting
the string-frame time coordinate $t_S$ as a function of $t_\s$ (see Fig.~\ref{stringtime}).
As the system collapses, the coordinate $t_S$ steadily ceases to evolve with increasing
$t_\s$, and reaches $t_S = 0$ only asymptotically at $t_\s = \infty$.  So, from the 
point of view of string frame, the system avoids reaching the singularity because
the dynamics are severely redshifted; the tachyon generates a smooth cutoff 
of cosmological time $t_S$.  To be certain, there is a cosmological Big Crunch at finite
time, but from the point of view of the underlying fundamental string, the physics is 
completely smooth.

In Einstein frame, the action takes the form
\be
S &=& \frac{1}{2\kappa^2} \int d^D X \sqrt{-\det G^E}
	\Bigl[
	R^E - \pp_\m \phi \pp^\m \phi
	-\frac{4}{\sqrt{D-2}} \,T\, \pp_\m \phi \pp^\m T
\xxx
&&
	-\frac{4}{D-2} \,(T^2 + D-2)\,	\pp_\m T \pp^\m T
	-\frac{2}{3\apr}\,e^{\frac{2\phi}{\sqrt{D-2}}}\, e^{\frac{2\,T^2}{D-2}} \, (D-26 - 24\,T^2)
	\Bigr] \ .
\label{neweffactEIN}
\ee
Expressed in this frame, the scale factor acquires a dependence on the dilaton gradient
$q = -\pp_{t_\s} \Phi$:
\be
a_E(t_\s,X^1) = \exp{ -\frac{1}{D-2}\left( \m^2 e^{\sqrt{2}\,\b\,(t_\s + X^1)} 
		- 2\,q\,t_\s \right) } \ .
\label{nullscale}
\ee
The FRW time coordinate evolves with $t_\s$ according to
\be
t_E(t_\s) = \int_1^{t_\s} \,d\xi \, \exp{ -\frac{1}{D-2} \left(2\,q\,\xi
	- e^{\sqrt{2}\,\b\,(X^1 + \xi)} \right) } + {\rm const.}
\label{EINtime}
\ee
Once again, this expression can be formally inverted to express $t_\s$ as a function of $t_E$
and $X^1$.  
By construction, the scale factor in Eqn.~\rr{nullscale}, 
and the tachyon and dilaton profiles in Eqn.~\rr{nullsolution}, are implicit 
particular solutions of the equations of motion.

\begin{figure}[htb]
\begin{center}
\includegraphics[width=3.5in,height=2.8in,angle=0]{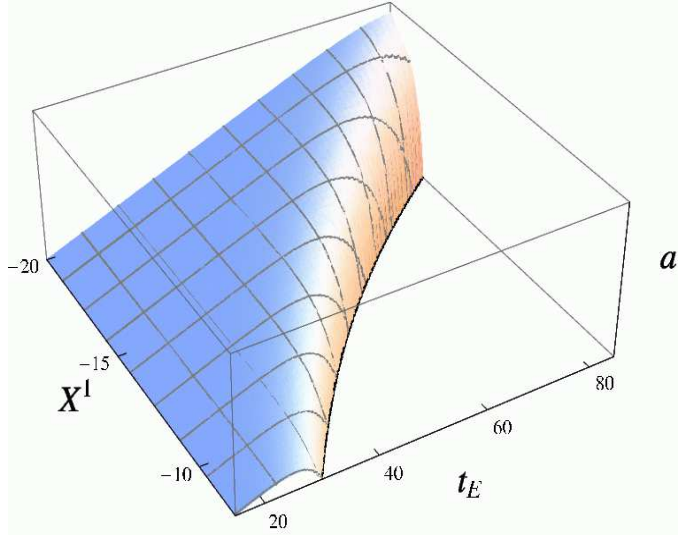}
\caption{The bubble of nothing in Einstein frame is a surface of 
asymptotically vanishing metric.
The scale factor initially grows linearly as a function of $t_E$, then collapses to a
near-singular region that expands outward in the negative $X^1$ direction
($D=1000$, $\m = 0.5$).  }
\label{bubble}
\end{center}
\end{figure}

Fig.~\ref{bubble} depicts the evolution of the scale factor in Einstein frame.  
When the tachyon is small, the universe 
evolves approximately linearly in $t_E$ for all $X^1$.  This is just a reflection of the fact that 
at zero tachyon the theory exhibits an equation of state that is precisely critical ($w = -(D-3)/(D-1)$).
As the tachyon increases in strength, the scale factor collapses, approaching a singularity 
asymptotically at $t_\s = \infty$.   To an observer at some fixed negative $X^1$, the region of
collapse appears to expand in the negative $X^1$ direction outward from the origin.
As the scale factor approaches zero, the cosmological time $t_E$ ceases to evolve as
a function of $t_\s$, and spacetime becomes frozen in a near-singularity.

\subsection{Toy models of the Big Bang}
As noted above, the beta function equations of the $2D$ worldsheet CFT stipulate 
that the dilaton gradient
$q = -\pp_{t_\s} \Phi$ satisfy
\be
q^2 = \left({\frac{D-26}{6\, \apr}}\right)^2 \ .
\ee
We can also study the second branch of this solution, where $q$ is negative definite. 
In this case, the dilaton evolves from weak coupling in the far past and rolls in the 
direction of strong coupling toward the future.  The linearized tachyon equation of motion
\rr{tachyonmarginality} is satisfied under the condition $\b = \frac{2\sqrt{2}}{  q \apr}$, 
so $\b$ is also 
negative definite.  Choosing this branch is therefore equivalent to invoking  
an overall time reversal on the the previous solution.  
The strength of the tachyon thus decreases with evolving time, reaching zero in the 
infinite future.  
(This general setup of a tachyonic Big Bang 
was studied in detail in \cite{Brandenberger:2007dh},
with a particular focus on the importance of fluctuations.)
Although this is essentially an extremely fine-tuned solution, let 
us briefly consider the picture that emerges in its own right.

\begin{figure}[htb]
\begin{center}
\includegraphics[width=2.7in,height=1.9in,angle=0]{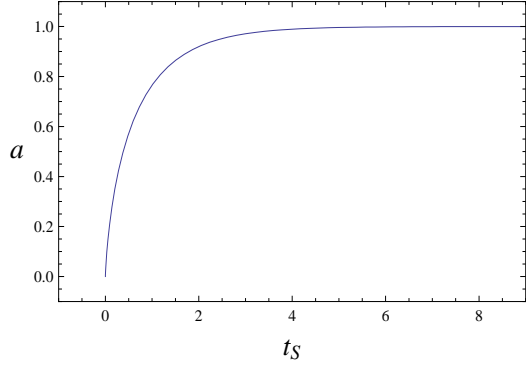}
\caption{A toy model of the Big Bang ($D=1000$, $\mu=0.5$, $X^1 = 0$).  The second branch
of the null tachyon solution causes the scale factor to emerge from a near-singularity and rapidly
accelerate, reaching a constant value (in string frame) when the tachyon shrinks to zero size.  }
\label{bang}
\end{center}
\end{figure}

Deep in the weakly coupled regime,  
the tachyon is large, and the FRW scale factor is correspondingly small.  As the tachyon
decreases in strength, the scale factor rapidly accelerates from a near-singularity.
(In Einstein frame, the evolution of the dilaton eventually takes over, and the scale factor contracts  
linearly as a function of $t_E$ for fixed $X^1$.)
This situation presents an interesting toy model of the Big Bang that can be studied 
at arbitrarily weak string coupling.  Fig.~\ref{bang} depicts the scale factor
as a function of $t_S$.   Deep in the past the scale factor is near 
zero and, as the tachyon passes below a critical value, the universe rapidly expands 
outward.  
 
When viewed as a function of $t_\s$, the temporal coordinate in string frame is 
also frozen deep in the past.  This is depicted in Fig.~\rr{frozen}, which displays a plot
of $t_S(t_\s)$.  The ``flow of time'' in string frame is thus generated as the tachyon
evolves toward zero.  

\begin{figure}[htb]
\begin{center}
\includegraphics[width=2.7in,height=1.9in,angle=0]{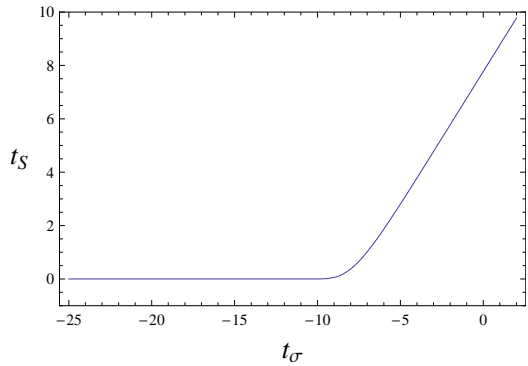}
\caption{In a tachyonic Big Bang, cosmological time
in string frame is frozen in the deep past,
as a function of sigma-model time $t_\s$. As the tachyon evolves toward 
zero, time begins to flow monotonically in the positive direction 
($D=1000$, $\m = 0.5$).  }
\label{frozen}
\end{center}
\end{figure}

\section{Solitonic solutions}
\label{solitons}
Part of the benefit of having a specific effective action in hand is the 
ability to study new solutions that are not used as input in deriving the action itself.
In this section we look for static solutions of the effective action that localize the
universe in spatial, rather than temporal directions.  

The general problem of using closed string tachyon condensation to localize or remove a 
spatial dimension has been studied in detail in 
\cite{paper0,paper2,moeller,BergmanRazamat,Green:2007tr}.
In \cite{paper2}, the tractability of the
null tachyon was used to derive {\it exact} solutions that dynamically remove spatial dimensions
from the theory.  The central charge deficit that arises in the classical worldsheet CFT from
the removal of these dimensions is made up by one-loop quantum corrections.  Because these 
solutions lie outside of the class of tachyon perturbations considered here \rr{solutions},
and because they are inherently quantum mechanical on the worldsheet, we will not
attempt to study them directly in the present context.

Instead, we adopt the strategy of \cite{BergmanRazamat}, which studied this problem 
directly at the level of effective actions.  The authors of \cite{BergmanRazamat}
were able to find toy models that localize spatial directions in the presence of 
general tachyon potentials.  We will follow this analysis in searching for analogous solutions
of the effective action computed above.  

A simple approach is to
introduce as an ansatz a spatially varying sigma model metric of the form:
\be
ds_\s^2 = (dX^1)^2 + \tilde a(X^1)^2\, \eta_{\m'\n'}\, dX^{\m'} dX^{\n'} \ ,
\qquad \m',\,\n' \in 0,2,3,\ldots,D-1 \ ,
\ee
with a warp factor $\tilde a(X^1)$ that depends only on $X^1$.
Following \cite{BergmanRazamat}, we observe that, in general, the dilaton will depend on
all of the embedding coordinates $X^\m $, though by the $(D-1)$-dimensional
Lorentz invariance we can move to a frame in which the 
dilaton depends only $X^1$ and $X^2$:
\be
\Phi = \Phi(X^1,X^2) \ .
\ee

The aim is to find codimension one solitons,
in which the tachyon depends solely on the spatial direction $X^1$. 
The nontrivial components of the Einstein equation (\ref{EINSTEIN}) in sigma
model frame yield the conditions (for general $f_1(T)$)
\be
\label{solein1}
0 & = & f_1\, \left[
		(D-2)(\at')^2 + \at\, \at'' - 2\, \at\, \at'\, \pp_1 \Phi
	\right]
	+ f_1'\,T'\, \at\, \at' \ ,
\\ &&\nn \\
\label{solein2}
0 & = & f_1\,\left(
		2\,\pp_1^2 \Phi - (D-1) \frac{\at''}{\at} \right)
	+f_1' \, \left( \frac{{T'}^2}{T} -  T'' \right)  \ ,
\\ &&\nn \\
\label{solein3}
0 & = & \frac{2\,f_1}{\at} \left( \at\, \pp_1\pp_2\Phi - \at'\, \pp_2\Phi \right)\ ,
\\ &&\nn \\
0 & = & - f_1 \left[
		(D-2)(\at')^2 + \at\,\at'' - 2 (\pp_2^2\Phi + \at\,\at'\,\pp_1\Phi)
	\right]  - f_1'\,T'\,\at\,\at'  \ .  
\label{solein4}
\ee
Eqns.~\rr{solein1} and \rr{solein4} combine to give the condition
\be
f_1\,\pp_2^2 \Phi = 0 \ .
\label{solcond1}
\ee
Eqn.~\rr{solein1} alone implies that $\pp_1\Phi$ can only depend on the $X^1$ coordinate
($f_1$ is a function of $T$ only, which we assume depends only on $X^1$), so
\be
\pp_2 \pp_1 \Phi = 0 \ .
\label{solcond2}
\ee
With the general condition that $f_1(T)$ is nonzero at finite $T$, we obtain the following
generic form for the dilaton:
\be
\Phi(X^1,X^2) = F(X^1) + Q\, X^2 \ ,
\ee
where $F(X^1)$ is some function of $X^1$ only, and $Q$ is a constant.
This same condition was derived for the codimension one soliton configurations
studied in \cite{BergmanRazamat}.
Eqn.~\rr{solein3} then reduces to
\be
Q\, \at'   = 0 \ ,
\ee
implying that either the sigma model metric is precisely flat, or the dilaton is 
independent of $X^2$.

Let us first consider solutions with a flat sigma model metric: 
\be
ds_\s^2 \equiv \eta_{\m\n}\, dX^\m dX^\n \ .
\ee
In this case, the remaining nontrivial component of the Einstein equation appears as
\be
2\,f_1\,F'' - f_1'\,T'' + \frac{f_1'}{T}\,{T'}^2 = 0 \ .
\ee
The dilaton and tachyon equations of motion take the form
\be
0 & = & f_1 \left( {F'}^2 + Q^2 \right)
	- f_1'\,F'\,T'
	+\frac{1}{4}\left( f_1'' + \frac{f_1'}{T}\right) {T'}^2
	+ \frac{1}{2} V_\s \ ,
\xxx
0 & = & \left(\frac{f_1'}{T} - f_1'' - f_1'''\,T \right) {T'}^2
	+ 4\,f_1'\,T (F'' - {F'}^2 - Q^2 )
\xxx
&&	+ 2(f_1' + f_1''\,T)(2\,F'\,T' - T'')
	- 2\,T\,\pp_T V_\s \ .
\ee
Employing the perturbative solution $f_1(T) \approx 1 - T^2$ and substituting the
tachyon potential into these equations yields the following conditions
on the transverse component of the dilaton:
\be
F' & = & \frac{1}{2\apr T'}\left( 4 T + \apr T'' \right) \ ,
\xxx
F'' & = & \frac{1}{(T^2-1)}\left( T T'' - T'^2 \right) \ .
\label{fconditions}
\ee
We also recover an explicit expression for the longitudinal dilaton gradient
in terms of the tachyon 
\be
Q^2  &=&  -\frac{1}{6(T^2-1)}\Bigl[
	6\, T'^2 - 6\,T\,T''
	+ \frac{1}{\apr}(T^2-1)(D-26)
\xxx
&&\kern+60pt
	+ \frac{3}{2\apr\, T'^2} (T^2-1)(4\,T + \apr\,T'')^2
	\Bigr] \ .
\ee

The conditions on the transverse dilaton $F(X^1)$ in Eqn.~\rr{fconditions}
impose the following differential equation for the tachyon profile:
\be
\frac{1}{(T^2-1)}(T'^2 - T\,T'')
	+ \frac{1}{2\apr\,T'^2}\left[
	4\,T'^2 - T''(4\,T + \apr\,T'')
	+ \apr\,T'\,T'' \right]  = 0\ .
\label{difftach}
\ee
This equation is satisfied exactly by the exponential profile
\be
T = \mu\, \exp{\b_1 X^1} \ .
\label{spaceliketachyon}
\ee
With this solution, the 
transverse dilaton is linear in $X^1$, with a gradient given by
\be
F' = \frac{2}{\apr\,\b_1} + \frac{\b_1}{2} \ .
\label{spacelikedilatongradient}
\ee
The longitudinal dilaton gradient takes the 
form\footnote{One can check that the tachyon profile satisfies the marginality condition, and
$Q^2 + F'^2 = - \frac{D-26}{6\apr}$.} 
\be
Q^2 = -\frac{\b_1^2}{4} - \frac{4}{\apr^2\b_1^2} - \frac{D-14}{6\,\apr} \ .
\label{Qgradient}
\ee
For real $\b_1$, $Q^2$ can only be nonnegative for $D \leq 2$.
(In fact, this conclusion can be reached using the 
full exponential form $f_1(T) = \exp{-T^2}$, or the cosh form 
$f_1(T) = 1 / \cosh(\sqrt{2}\,T)$, in the above equations.)
If a timelike direction is present at all, the {\it only} consistent solution exists 
in $D=2$, with $Q = 0$.  The resulting system is described by 
an exponential tachyon profile with a spacelike 
linear dilaton, both varying in the $X^1$ direction.
The dilaton gradient and tachyon profile are determined by
\be
F' = \b_1 = \pm \frac{2}{\sqrt{\apr}} \ .
\ee
Therefore, the only static solution consistent with a single 
exponential tachyon and an exactly flat
sigma model metric lives strictly in $D=2$.

Of course, this restriction is lifted if the dilaton is allowed to vary in the 
timelike direction (so the configuration is no longer static).  
For example, consider the form
\be
\Phi(X^0,X^1) = F(X^1) - q\, X^0 \ .
\ee
In this case, with a tachyon profile of the form in Eqn.~\rr{spaceliketachyon},
the dilaton is again linear along the $X^1$ direction, with gradient given by 
Eqn.~\rr{spacelikedilatongradient}.  However, the gradient in the timelike direction
is given by \rr{Qgradient}, with an overall sign change: $q^2 = -Q^2$.  This is 
positive definite for all $D \geq 2$.  The special case in which $F' = q$ 
corresponds to spacelike Liouville theory with a null linear dilaton in the critical
dimension $D=26$.

It is instructive to look for other static tachyon solutions.
One method is to solve \rr{difftach} order-by-order in $\apr$.  To
$O(\apr^3)$, one obtains
\be
T(X^1) &=& \mu\, e^{\b_1 X^1}
	+ \apr\, e^{\b_1 X^1}\,(c_1 + c_2 X^1 )
	+ \apr^2\, e^{\b_1 X^1}\, \Bigl( c_3 + c_4 X^1 + \frac{1}{2\m} c_2^2 (X^1)^2 \Bigr)
\xxx
&&
\kern-50pt
	+ \apr^3\, e^{\b_1 X^1}\, \Bigl( c_5 + c_6 X^1 
	+ \frac{1}{\m} \bigl(
		c_2\,c_4
		- \frac{1}{2\m} c_1\,c_2^2 \bigr) \,(X^1)^2 
	+ \frac{1}{6\m^2}c_2^3 (X^1)^3 \Bigr) 
	+ O(\apr^4)\ ,
\label{tachexpsolution}
\ee
where $c_n$ are free constants.  It is easy to find a set of constants $c_n$ such that the tachyon
exhibits a ``lump'' configuration over some intermediate range in $X^1$ (and the square of the 
longitudinal dilaton gradient is positive definite).

\begin{figure}[htb]
\begin{center}
\subfigure[Tachyon]{
\includegraphics[width=2.5in,height=1.8in,angle=0]{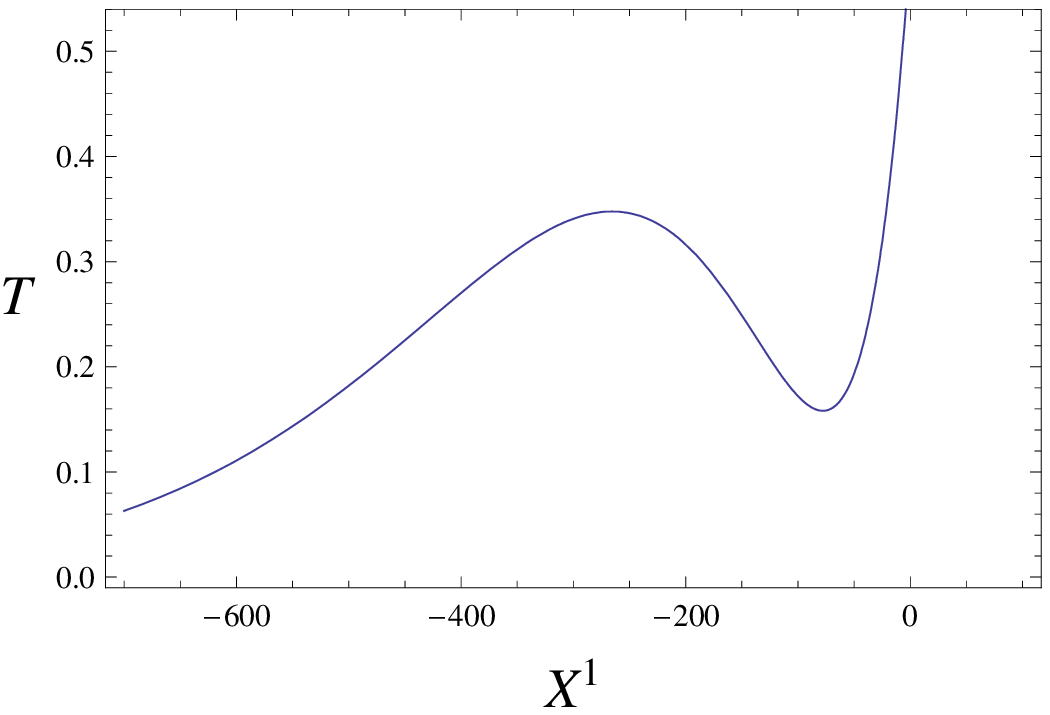}}
\hspace{8mm}
\subfigure[Einstein-frame warp factor]{
\includegraphics[width=2.5in,height=1.8in,angle=0]{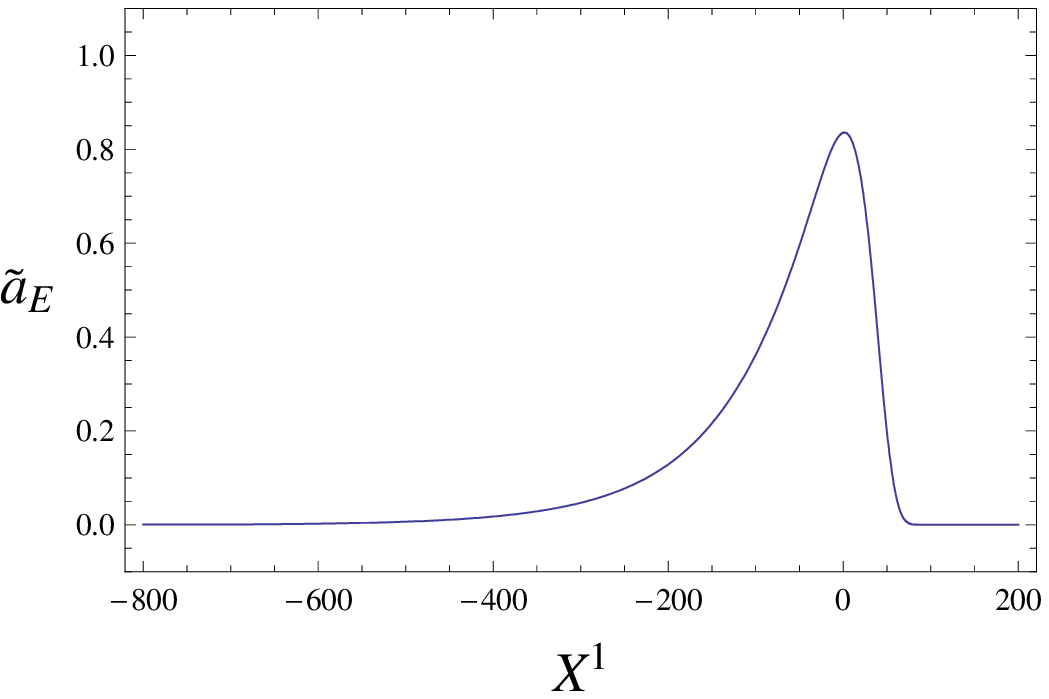}}
\caption{A lump configuration of the spatially-varying tachyon (panel (a)) with
flat sigma model metric.  The
asymptotic regions are dominated by an overall exponential prefactor in the solution that
is present to all orders in $\apr$.  The Einstein-frame metric (panel (b)) exhibits a 
solitonic configuration of localized spacetime.  }
\label{lump}
\end{center}
\end{figure}

The exponential prefactor $\exp{\b_1 X^1}$
is present in the solution \rr{tachexpsolution} to all orders in $\apr$, multiplying
terms that are universally polynomial in $X^1$.  The prefactor therefore 
dominates in the asymptotic regions, and   
the tachyon is forced to vanish at $X^1 = \pm \infty$, depending on the sign of $\b_1$.
An example lump profile is depicted in Fig.~\ref{lump} (panel (a)).   In string frame, the metric
exhibits a configuration that is finite over a semi-infinite region, 
with a localized ``pseudo-soliton'' in some separate region.
The resulting picture is a semi-infinite universe in $D$ dimensions, with an effectively  
lower-dimensional neighboring parallel universe.
In Einstein frame, the semi-infinite region of finite metric can be removed by arranging the 
transverse dilaton to increase in the appropriate direction.  This is displayed in 
Fig.~\ref{lump} (panel (b)), with $f_1(T) = \exp{-T^2}$. 

Of course, one should keep in mind that the effective action itself is not an exact description
of string theory, and higher-order effects are likely to become important in the absence of
some special mechanism (as with the null tachyon).  
Even so, the leading-order behavior of non-exact solutions can often serve as a 
useful qualitative guide in determining the types of solutions that are possible.

In addition, one should also consider solutions for which the sigma model metric
is nonflat.  In these cases, the dilaton must have a vanishing 
longitudinal gradient.\footnote{A recent paper \cite{moeller} has provided 
evidence for the existence of a 
codimension one soliton in closed string field theory.  In that analysis, the 
dilaton did not vary longitudinally along the soliton, though the more general
case was not considered directly.}  
To study the ability of a solitonic configuration to localize
the universe along a spatial direction, it is most natural to work in Einstein frame,
and a straightforward approach is to search for numerical solutions to the equations of
motion that exhibit the qualitative properties of interest.    
If the dilaton and tachyon are arranged to increase in opposite coordinate directions 
(like all of the examples studied above), the warp factor in Einstein frame $\tilde a_E$
will naturally acquire local support in a region where both the dilaton and tachyon are small.

\begin{figure}[htb]
\begin{center}
\includegraphics[width=2.5in,height=1.8in,angle=0]{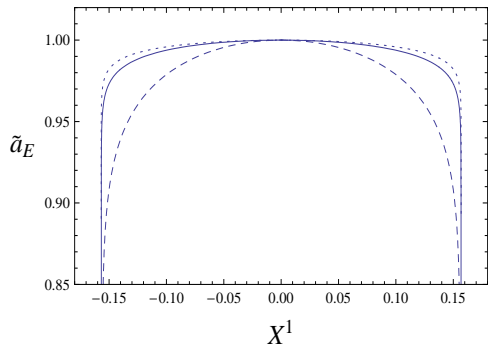}
\caption{An example of a solitonic tachyon configuration that localizes spacetime in
a spatial direction, at $D=30$ (dashed), $D=100$ (solid) and $D=150$ (dotted).  Increasing $D$
brings the scale factor closer to a constant over the finite region of the solution.  The
overall scale of the localized dimension is roughly constant for a given set of 
boundary conditions. }
\label{solitonsplot}
\end{center}
\end{figure}

\begin{figure}[htb]
\begin{center}
\subfigure[Dilaton $\phi$]{
\includegraphics[width=2.5in,height=1.8in,angle=0]{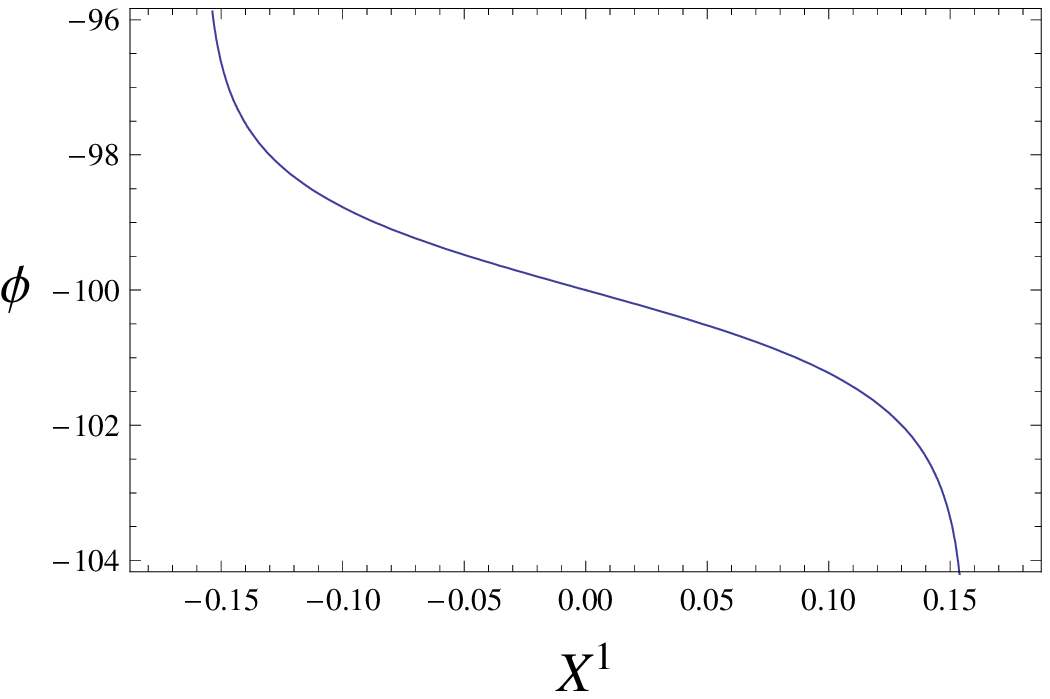}}
\hspace{8mm}
\subfigure[Tachyon $T$]{
\includegraphics[width=2.5in,height=1.8in,angle=0]{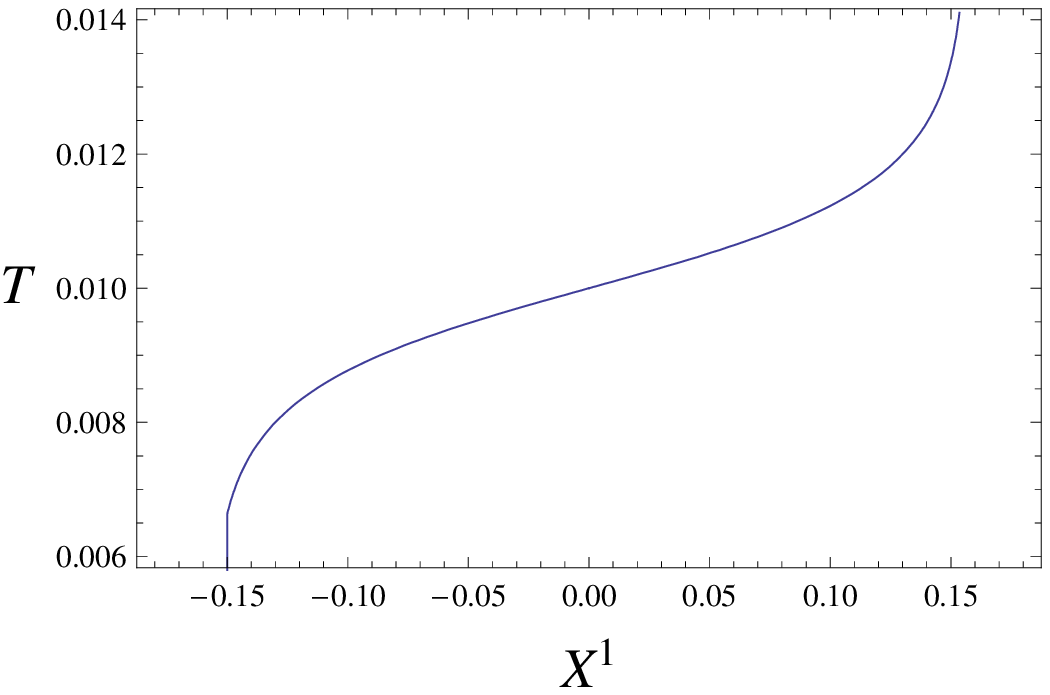}}
\caption{Behavior of the dilaton and tachyon in a solitonic configuration ($D=100$).  
With a given set of boundary conditions, the numerical solutions for the dilaton
and tachyon do not vary visibly with varying dimension. }
\label{solitonsplot2}
\end{center}
\end{figure}

One such set of numerical solutions is displayed for various $D$ 
in Fig.~\ref{solitonsplot}.  In these solutions, the dilaton increases
in the negative $X^1$ direction, while the tachyon magnitude increases in the positive $X^1$ direction.
This is depicted in Fig.~\ref{solitonsplot2}.
Along the negative $X^1$ axis, the growth of the dilaton drives the warp
factor toward zero size, while along the positive $X^1$ direction the growth of the tachyon drives 
$\at_E$ toward zero as well.   The scale factor can exist at finite size in the region in between.
Taken at face value, the overall scale of the confined dimension is essentially a 
function of initial conditions. 
For a given set of initial conditions, the dilaton and tachyon solutions do not vary 
considerably with shifting $D$ (see Fig.~\ref{solitonsplot2}).
Of course, the string theory can become strongly coupled deep in the region of negative 
$X^1$, so the solution is subject to corrections there.  Furthermore, $\apr$ corrections 
can become important in regions that exhibit singular (or near-singular) behavior.  
Again, these types of solutions should therefore be viewed only as toy models.

\section{Summary and conclusions}
\label{conclusions}
We have seen that there is a unique two-derivative effective action that perturbatively 
supports the full class of solutions in Eqn.~\rr{solutions}.  
Finding this effective action amounts to finding a 
perturbative expansion for the function $f_1(T)$, defined 
as a prefactor of the Einstein-Hilbert term in sigma-model frame.  Demanding that the general
action support the linearized solutions in Eqn.~\rr{solutions} allowed us to solve for $f_1(T)$ to 
quadratic order.  

Taken at face value, the solutions of interest generically encounter singularities as the
tachyon evolves to become of order one.  This confirms the general expectations provided
by \cite{Yangbigcrunch}.  At $T \approx 1$, however, we expect these solutions
to become subject to higher-order corrections.  
We saw that the two-derivative effective action
supports the null tachyon for any $f_1(T)$, and possible modifications
from higher-order effects are restricted to affect the spacetime solution 
only through higher-order constraints to the function $f_1(T)$.  

Considering the effective action for completely general $f_1(T)$, we showed that singularities
in the string-frame action arise either from singularities of the metric itself, or from 
prefactors of the matter kinetic terms that arise from the Weyl transformation
in Eqn.~\rr{conformaltform}.   If we demand that the latter singularities are absent, and that
the gravity sector remains unitary for all $T$, then the curvature singularities of the null
tachyon system are resolved for all finite values of the tachyon itself.  
The resolution of these singularities suggests an interesting description of
how cosmological time can be initiated or terminated in string theory.  We expect there to 
be many applications of this sort of mechanism in more general cosmological models.

Of course, we should leave open the possibility that the framework of the spacetime effective
action is simply insufficient to capture the physics of bulk tachyon condensation, and that
a careful accounting of higher-order corrections to the action will reveal unavoidable
singularities arising at finite values of the tachyon, even for exact solutions.  
It would certainly be interesting to try to constrain possible higher-order corrections to the
effective action directly.  The general expectation is that higher-order 
effects in conformal perturbation theory
will amount to corrections to $f_1(T)$ beyond quadratic order.  There is not a unique
function that satisfies the conditions on $f_1(T)$ given in Section \ref{resolved},
but a test to determine whether corrections beyond $O(T^2)$ can resolve cosmological singularities 
is to see whether higher-order terms divert the function from crossing zero
at $T=1$.  A promising hint of this would be that the next term in the series 
is positive definite for all $T$.  
In other words, one might expect general higher-order effects to yield a 
correction to $f_1(T)$ of the form
\be
f_1(T) = 1 - T^2 + c\,T^n + O(T^{n+1}) \ ,
\ee
where $c$ is positive definite and $n$ is an even integer.  
It would clearly be valuable to test this prediction by direct methods.

\section*{Acknowledgments}
The author thanks Oren Bergman and Simeon Hellerman for helpful and interesting 
discussions, and for reading a draft
of the manuscript; he thanks David Kutasov, Martin Schnabl and Sav Sethi for useful 
discussions.   I.S.~is supported by the Marvin L.~Goldberger Membership
at the Institute for Advanced Study, and by U.S.~National Science Foundation grant PHY-0503584. 


\bibliographystyle{utcaps}
\bibliography{effact}

\providecommand{\href}[2]{#2}\begingroup\raggedright\begin{thebibliography}{10}

\bibitem{Sen:2004nf}
A.~Sen, ``{Tachyon dynamics in open string theory},'' {\em Int. J. Mod. Phys.}
  {\bf A20} (2005) 5513--5656,
\href{http://www.arXiv.org/abs/hep-th/0410103}{{\tt hep-th/0410103}}.

\bibitem{Taylor:2003gn}
W.~Taylor and B.~Zwiebach, ``{D-branes, tachyons, and string field theory},''
\href{http://www.arXiv.org/abs/hep-th/0311017}{{\tt hep-th/0311017}}.

\bibitem{DeSmet:2001af}
P.-J. De~Smet, ``{Tachyon condensation: Calculations in string field theory},''
\href{http://www.arXiv.org/abs/hep-th/0109182}{{\tt hep-th/0109182}}.

\bibitem{Ohmori:2001am}
K.~Ohmori, ``{A review on tachyon condensation in open string field
  theories},''
\href{http://www.arXiv.org/abs/hep-th/0102085}{{\tt hep-th/0102085}}.

\bibitem{martinsimeon}
S.~Hellerman and M.~Schnabl, ``{Light-like tachyon condensation in Open String
  Field Theory},''
\href{http://www.arXiv.org/abs/0803.1184}{{\tt 0803.1184}}.

\bibitem{local1}
A.~Adams, J.~Polchinski, and E.~Silverstein, ``{Don't panic! Closed string
  tachyons in ALE space-times},'' {\em JHEP} {\bf 10} (2001) 029,
\href{http://www.arXiv.org/abs/hep-th/0108075}{{\tt hep-th/0108075}}.

\bibitem{local2}
C.~Vafa, ``{Mirror symmetry and closed string tachyon condensation},''
\href{http://www.arXiv.org/abs/hep-th/0111051}{{\tt hep-th/0111051}}.

\bibitem{local3}
J.~A. Harvey, D.~Kutasov, E.~J. Martinec, and G.~W. Moore, ``{Localized
  tachyons and RG flows},''
\href{http://www.arXiv.org/abs/hep-th/0111154}{{\tt hep-th/0111154}}.

\bibitem{local4}
A.~Dabholkar, ``{Tachyon condensation and black hole entropy},'' {\em Phys.
  Rev. Lett.} {\bf 88} (2002) 091301,
\href{http://www.arXiv.org/abs/hep-th/0111004}{{\tt hep-th/0111004}}.

\bibitem{local5}
R.~Gregory and J.~A. Harvey, ``{Spacetime decay of cones at strong coupling},''
  {\em Class. Quant. Grav.} {\bf 20} (2003) L231--L238,
\href{http://www.arXiv.org/abs/hep-th/0306146}{{\tt hep-th/0306146}}.

\bibitem{local6}
M.~Headrick, ``{Decay of C/Z(n): Exact supergravity solutions},'' {\em JHEP}
  {\bf 03} (2004) 025,
\href{http://www.arXiv.org/abs/hep-th/0312213}{{\tt hep-th/0312213}}.

\bibitem{local7}
A.~Adams, X.~Liu, J.~McGreevy, A.~Saltman, and E.~Silverstein, ``{Things fall
  apart: Topology change from winding tachyons},'' {\em JHEP} {\bf 10} (2005)
  033,
\href{http://www.arXiv.org/abs/hep-th/0502021}{{\tt hep-th/0502021}}.

\bibitem{local8}
T.~Suyama, ``{Tachyons in compact spaces},'' {\em JHEP} {\bf 05} (2005) 065,
\href{http://www.arXiv.org/abs/hep-th/0503073}{{\tt hep-th/0503073}}.

\bibitem{local9}
Y.~Okawa and B.~Zwiebach, ``{Twisted tachyon condensation in closed string
  field theory},'' {\em JHEP} {\bf 03} (2004) 056,
\href{http://www.arXiv.org/abs/hep-th/0403051}{{\tt hep-th/0403051}}.

\bibitem{paper0}
S.~Hellerman and X.~Liu, ``{Dynamical dimension change in supercritical string
  theory},''
\href{http://www.arXiv.org/abs/hep-th/0409071}{{\tt hep-th/0409071}}.

\bibitem{paper1}
S.~Hellerman and I.~Swanson, ``{Cosmological solutions of supercritical string
  theory},''
\href{http://www.arXiv.org/abs/hep-th/0611317}{{\tt hep-th/0611317}}.

\bibitem{paper2}
S.~Hellerman and I.~Swanson, ``{Dimension-changing exact solutions of string
  theory},'' {\em JHEP} {\bf 09} (2007) 096,
\href{http://www.arXiv.org/abs/hep-th/0612051}{{\tt hep-th/0612051}}.

\bibitem{paper4}
S.~Hellerman and I.~Swanson, ``{Charting the landscape of supercritical string
  theory},'' {\em Phys. Rev. Lett.} {\bf 99} (2007) 171601,
\href{http://www.arXiv.org/abs/0705.0980}{{\tt 0705.0980}}.

\bibitem{csft1}
B.~Zwiebach, ``{Closed string field theory: Quantum action and the B-V master
  equation},'' {\em Nucl. Phys.} {\bf B390} (1993) 33--152,
\href{http://www.arXiv.org/abs/hep-th/9206084}{{\tt hep-th/9206084}}.

\bibitem{csft2}
M.~Saadi and B.~Zwiebach, ``{Closed String Field Theory from Polyhedra},'' {\em
  Ann. Phys.} {\bf 192} (1989)
213.

\bibitem{csft3}
T.~Kugo, H.~Kunitomo, and K.~Suehiro, ``{Nonpolynomial Closed String Field
  Theory},'' {\em Phys. Lett.} {\bf B226} (1989)
48.

\bibitem{csft4}
T.~Kugo and K.~Suehiro, ``{Nonpolynomial closed string field theory: Action and
  its gauge invariance},'' {\em Nucl. Phys.} {\bf B337} (1990)
434--466.

\bibitem{csft5}
M.~Kaku, ``{Geometric derivation of string field theory from first principles:
  Closed strings and modular invariance},'' {\em Phys. Rev.} {\bf D38} (1988)
3052.

\bibitem{csft6}
M.~Kaku and J.~D. Lykken, ``{Modular invariant closed string field theory},''
  {\em Phys. Rev.} {\bf D38} (1988)
3067.

\bibitem{csftvac1}
V.~A. Kostelecky and S.~Samuel, ``{Collective physics in the closed bosonic
  string},'' {\em Phys. Rev.} {\bf D42} (1990)
1289--1292.

\bibitem{csftvac2}
A.~Belopolsky and B.~Zwiebach, ``{Off-shell closed string amplitudes: Towards a
  computation of the tachyon potential},'' {\em Nucl. Phys.} {\bf B442} (1995)
  494--532,
\href{http://www.arXiv.org/abs/hep-th/9409015}{{\tt hep-th/9409015}}.

\bibitem{csftvac3}
A.~Belopolsky, ``{Effective Tachyonic potential in closed string field
  theory},'' {\em Nucl. Phys.} {\bf B448} (1995) 245--276,
\href{http://www.arXiv.org/abs/hep-th/9412106}{{\tt hep-th/9412106}}.

\bibitem{csftvac4}
N.~Moeller, ``{Closed bosonic string field theory at quartic order},'' {\em
  JHEP} {\bf 11} (2004) 018,
\href{http://www.arXiv.org/abs/hep-th/0408067}{{\tt hep-th/0408067}}.

\bibitem{Moeller:2006cw}
N.~Moeller, ``{Closed bosonic string field theory at quintic order: Five-
  tachyon contact term and dilaton theorem},'' {\em JHEP} {\bf 03} (2007) 043,
\href{http://www.arXiv.org/abs/hep-th/0609209}{{\tt hep-th/0609209}}.

\bibitem{Moeller:2007mu}
N.~Moeller, ``{Closed bosonic string field theory at quintic order. II:
  Marginal deformations and effective potential},'' {\em JHEP} {\bf 09} (2007)
  118,
\href{http://www.arXiv.org/abs/0705.2102}{{\tt 0705.2102}}.

\bibitem{Bergman:2004st}
O.~Bergman and S.~S. Razamat, ``{On the CSFT approach to localized closed
  string tachyons},'' {\em JHEP} {\bf 01} (2005) 014,
\href{http://www.arXiv.org/abs/hep-th/0410046}{{\tt hep-th/0410046}}.

\bibitem{csftvac5}
H.~Yang and B.~Zwiebach, ``{A closed string tachyon vacuum?},'' {\em JHEP} {\bf
  09} (2005) 054,
\href{http://www.arXiv.org/abs/hep-th/0506077}{{\tt hep-th/0506077}}.

\bibitem{moeller}
N.~Moeller, ``{A tachyon lump in closed string field theory},''
\href{http://www.arXiv.org/abs/0804.0697}{{\tt 0804.0697}}.

\bibitem{Bergman:2005qf}
O.~Bergman and S.~Hirano, ``{Semi-localized instability of the Kaluza-Klein
  linear dilaton vacuum},'' {\em Nucl. Phys.} {\bf B744} (2006) 136--155,
\href{http://www.arXiv.org/abs/hep-th/0510076}{{\tt hep-th/0510076}}.

\bibitem{Aharony:2006ra}
O.~Aharony and E.~Silverstein, ``{Supercritical stability, transitions and
  (pseudo)tachyons},'' {\em Phys. Rev.} {\bf D75} (2007) 046003,
\href{http://www.arXiv.org/abs/hep-th/0612031}{{\tt hep-th/0612031}}.

\bibitem{paper3}
S.~Hellerman and I.~Swanson, ``{Cosmological unification of string theories},''
\href{http://www.arXiv.org/abs/hep-th/0612116}{{\tt hep-th/0612116}}.

\bibitem{paper5}
S.~Hellerman and I.~Swanson, ``{Supercritical N = 2 string theory},''
\href{http://www.arXiv.org/abs/0709.2166}{{\tt 0709.2166}}.

\bibitem{paper6}
S.~Hellerman and I.~Swanson, ``{A stable vacuum of the tachyonic E8 string},''
\href{http://www.arXiv.org/abs/0710.1628}{{\tt 0710.1628}}.

\bibitem{wittenbubble}
E.~Witten, ``{Instability of the Kaluza-Klein Vacuum},'' {\em Nucl. Phys.} {\bf
  B195} (1982)
481.

\bibitem{BergmanRazamat}
O.~Bergman and S.~S. Razamat, ``{Toy models for closed string tachyon
  solitons},'' {\em JHEP} {\bf 11} (2006) 063,
\href{http://www.arXiv.org/abs/hep-th/0607037}{{\tt hep-th/0607037}}.

\bibitem{Yangbigcrunch}
H.~Yang and B.~Zwiebach, ``{Rolling closed string tachyons and the big
  crunch},'' {\em JHEP} {\bf 08} (2005) 046,
\href{http://www.arXiv.org/abs/hep-th/0506076}{{\tt hep-th/0506076}}.

\bibitem{Kutasov:2003er}
D.~Kutasov and V.~Niarchos, ``{Tachyon effective actions in open string
  theory},'' {\em Nucl. Phys.} {\bf B666} (2003) 56--70,
\href{http://www.arXiv.org/abs/hep-th/0304045}{{\tt hep-th/0304045}}.

\bibitem{Brandenberger:2007dh}
R.~H. Brandenberger, A.~R. Frey, and S.~Kanno, ``{Emergence of Fluctuations
  from a Tachyonic Big Bang},'' {\em Phys. Rev.} {\bf D76} (2007) 083524,
\href{http://www.arXiv.org/abs/0706.1104}{{\tt 0706.1104}}.

\bibitem{Green:2007tr}
D.~Green, A.~Lawrence, J.~McGreevy, D.~R. Morrison, and E.~Silverstein,
  ``{Dimensional Duality},'' {\em Phys. Rev.} {\bf D76} (2007) 066004,
\href{http://www.arXiv.org/abs/0705.0550}{{\tt 0705.0550}}.

\end{thebibliography}\endgroup

\end{document}